\documentclass{aa}  
\usepackage{graphicx}
\usepackage{txfonts}
\usepackage{float}
\begin{document} 

   \title{Solar photospheric spectrum microvariability }

   \subtitle{I. Theoretical searches for proxies of radial-velocity jittering}
  
   \author{Dainis Dravins
          \inst{1} and
       Hans-G\"{u}nter Ludwig
           \inst{2}
}
%
%
\institute{Lund Observatory, Division of Astrophysics, Department of Physics, Lund University, SE-22100 Lund, Sweden\\
              \email{dainis@astro.lu.se}
\and
     Zentrum f\"{u}r Astronomie der Universit\"{a}t Heidelberg, Landessternwarte, K\"{o}nigstuhl, DE--69117 Heidelberg, Germany\\
              \email{hludwig@lsw.uni-heidelberg.de}
             }
 
\date{Received June 9, 2023; accepted August 18, 2023}

 
\abstract
   {Extreme precision radial-velocity spectrometers enable extreme precision stellar spectroscopy.  Searches for low-mass exoplanets around solar-type stars are limited by various types of physical variability in stellar spectra, such as the short-term jittering of apparent radial velocities on levels of $\sim$2~m\,s$^{-1}$.}
   {To understand physical origins of radial-velocity jittering, the solar spectrum is assembled, as far as possible, from basic principles.  Solar surface convection is modeled with time-dependent 3D hydrodynamics, followed by the computation of high-resolution spectra during numerous instances of the simulation sequence.  The behavior of different classes of photospheric spectral lines is monitored throughout the simulations to identify commonalities or differences between different classes of lines: weak or strong, neutral or ionized, high- or low-excitation, atomic or molecular. }
   {Synthetic spectra are examined.  With a wavelength sampling $\lambda$/$\Delta\lambda$ $\sim$1,000,000, the changing shapes and wavelength shifts of unblended and representative \ion{Fe}{i} and \ion{Fe}{ii} lines are followed during the simulation sequences.  The radial-velocity jittering over the small simulation area typically amounts to $\pm$\,150~m\,s$^{-1}$, scaling to $\sim$2~m\,s$^{-1}$ for the full solar disk.  Flickering within the G-band region and in photometric indices of the Str{\"o}mgren $uvby$ system are also measured, and synthetic G-band spectra from magnetic regions discussed. }
   {Most photospheric lines vary in phase but with different amplitudes among different classes of lines.  Amplitudes of radial-velocity excursions are greater for stronger and for ionized lines, decreasing at longer wavelengths.  By matching precisely measured radial velocities to such characteristic patterns should enable to remove a significant component of the stellar `noise' originating in granulation. }
   {The granulation-induced amplitudes in full-disk sunlight amount to $\sim$2~m\,s$^{-1}$, with the differences between various line-groups an order of magnitude less.  To mitigate this jittering, a matched filter must recognize dissimilar lineshifts among classes of diverse spectral lines with precisions $\sim$10~cm\,s$^{-1}$ for each line group.  To verify the modeling toward such a filter, predictions of center-to-limb dependences of jittering amplitudes for different classes of lines are presented, testable with spatially resolving solar telescopes connected to existing radial-velocity instruments. }


\titlerunning{Solar spectrum microvariability. I}
\authorrunning{D. Dravins \& H.-G.Ludwig}
\maketitle

\section{Introduction}

Extreme precision stellar spectroscopy is now enabled at several observatories, using extreme precision radial-velocity spectrometers designed for a wavelength stability corresponding to Doppler shifts of $\sim$1~m\,s$^{-1}$ or less.  Such are operating at several large and very large telescopes, e.g., CARMENES on Calar Alto \citep{quirrenbachetal14, quirrenbachetal18}, ESPRESSO at the ESO VLT on Paranal \citep{pepeetal14, pepeetal21}; EXPRES at the Lowell Discovery Telescope in Arizona \citep{jurgensonetal16, blackmanetal20}; HARPS \citep{mayoretal03, coffinetetal19} and NIRPS \citep{bouchyetal17} at ESO on La Silla; HARPS-North and the forthcoming HARPS-3 on La Palma \citep{cosentinoetal12, thompsonetal16}, HPF on the Hobby-Ebery Telescope in Texas \citep{kanodiaetal21,mahadevanetal14}; iLocater on LBT in Arizona \citep{crassetal22}; KPF, the Keck Planet Finder on Maunakea \citep{gibsonetal16}; MAROON-X at Gemini-North on Maunakea \citep{seifahrtetal18, seifahrtetal22}; NEID at WYIN on Kitt Peak \citep{schwabetal16}; SPIRou at CFHT on Maunakea \citep{donatietal20, hobsonetal21}, Veloce Rosso at AAT in Australia \citep{gilbertetal18}, and a few others.  Such performance requirements also are in the design criteria of future instruments for extremely large telescopes \citep{halversonetal16, marconietal22, plavchanetal19, szentgyorgyietal18}.  

The primary task for these spectrometers is to search for periodic modulations of stellar radial velocities induced by orbiting exoplanets.  The detection of planets with successively smaller mass requires to recognize successively smaller velocity amplitudes: for an Earth-mass planet orbiting a solar-mass star in a one-year orbit, the expected signal amounts to at most only 0.1 m\,s$^{-1}$ \citep[e.g.,][]{halletal18}  The precision of current and foreseen instruments now actually begins to approach these levels, soon enabling such very challenging observations.  However, radial-velocity fluctuations of much greater amplitude are contributed by stellar phenomena such as convective motions in granulation and supergranulation, p-mode oscillations, modulation by dark starspots and bright magnetically active regions carried in and out of sight by stellar rotation; changes in surface structure during long-term magnetic activity cycles, and other.  The requirement of mitigating effects of stellar activity for radial-velocity measurements was realized already long ago \citep[e.g.,][]{dravins89} and to segregate such, often irregular, effects from the more regular modulation induced by planets, comprehensive data analysis methods have been developed to scrutinize spectra recorded with current instruments.  These typically include the calculation of cross-correlation or other statistical functions to find the wavelength displacement relative to a template, with various adjustments applied using the asymmetries or widths of such functions.  The level of magnetic activity can be estimated from simultaneously measured chromospheric emission, usually in the cores of the \ion{Ca}{ii}~H\,{\&}\,K-lines but also feasible from statistics of Zeeman broadening \citep{lienhardetal23}.  Extensive comparisons between different methods advocated by different groups have been carried out, testing software routines on data with both a synthetic planetary signal injected, and for sequences of pure stellar observations \citep{dumusque16, dumusqueetal17, haraford23, zhaoetal22}.

While such routines do succeed in eliminating much of the non-planetary signals, even the currently best modeling of stellar variability is unable to extract planetary signals with amplitudes much less than 1 or 2 m\,s$^{-1}$.  The limitations are thus no longer instrumental or technical but lie in understanding the complexities of atmospheric dynamics and spectral line formation, manifest as a jitter of the apparent radial velocity and flickering of the photometric irradiance.  As concluded in reviews of the field \citep{fischeretal16}, and in reports from working groups for precision radial velocities  and exoplanet detection  \citep{crassetal21, rackhametal23}, some significantly different approach will be required in order to reach precisions of 10~cm\,s$^{-1}$ or better, as ultimately required to find an exoEarth.

The aim of this paper is not to further refine methods in analyzing observed spectra but instead to start to try to understand the physical origins of radial-velocity jittering in solar and stellar atmospheres.  To reach challenging goals of precisions in the cm\,s$^{-1}$ range, it will probably not suffice to apply statistical analyses on observed data but to somehow exploit the information content of the many thousands of different spectral lines in solar-type stars.

We examine time sequences of synthetic solar spectra, as computed from hydrodynamic simulations of the solar photosphere.  In principle, this amounts to watching movies of high-resolution solar spectral atlases, searching for systematic patterns of variability, in particular such that correlate with excursions in apparent radial velocity.  The paper is organized as follows: Section 2 describes the background, the aims, the feasibility constraints, and places the velocity jittering over the simulation areas in the context of the full solar disk.  Section 3 describes how the overall synthetic spectra are obtained, and Section 4 how the properties of their individual spectral lines are measured.  Section  5 analyzes the wavelength jittering of groups of selected \ion{Fe}{i} and \ion{Fe}{ii} lines.  Section 6 examines the behavior of the molecular G-band and surrounding atomic lines, including a discussion of their formation physics and effects by magnetic fields. Section 7 evaluates the fluctuations in color indices from broader-band $uvby$ Str{\"o}mgren photometry.  Section 8 documents searches for such spectral-line parameters that failed to yield significant correlations with excursions in radial velocity.  The concluding Section 9 outlines how it should be possible to diminish the jittering signal arising from surface convection in searches for low-mass exoplanets.  The subsequent Paper II \citep{dravinsludwig24} will evaluate signatures of microvariabilty in observations of the Sun seen as a star, also guided by the conclusions from this Paper I.

\begin{figure*}
 \centering
 \includegraphics[width=18cm]{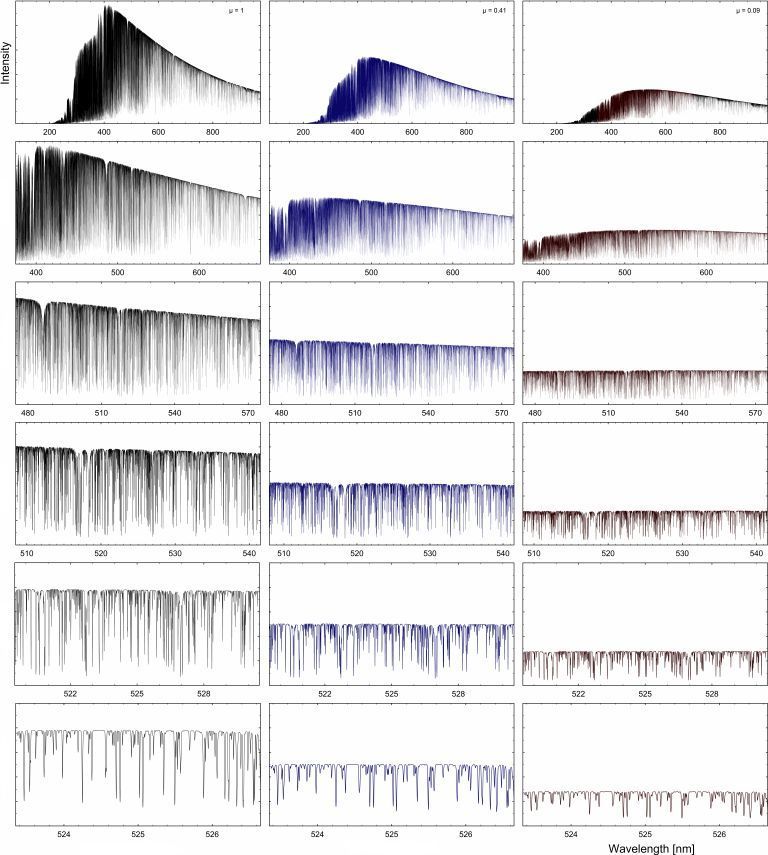}
    \caption{Synthetic solar spectra from the CO\,$^5$BOLD N59 model simulation, centered on 525 nm; seen over gradually narrower wavelength ranges, decreasing by a factor of three for each row from top down.  Left column: Disk center, $\mu$ = cos\,$\theta$ = 1, middle: $\mu$ = 0.41,  right: $\mu$ = 0.09.  The intensity scale is common for all frames, i.a., illustrating how the limb darkening is more pronounced at shorter wavelengths. }  
    \label{fig:full_spectrum}
\end{figure*}

\section{Synthesizing spectra from line formation physics}

A copious amount of work has been invested in analyzing and dissecting observed spectra, using a multitude of statistical and other functions, but appears to hit a boundary on the m\,s$^{-1}$ level.  Instead of scrutinizing observed spectra from `top down', we now try to assemble and construct not-yet-observed stellar spectra from `bottom up', as far as possible from basic principles.  By numerically simulating stellar atmospheric structure and dynamics `ab initio' for both quiet and magnetically active regions, and knowing the physical conditions in the model, signatures from dynamic stellar surface inhomogeneities can be traced forward into the complete synthetic spectrum.  Different classes of stellar surface features (hot or cold, magnetic or not) affect different classes of spectral lines (weak or strong, neutral or ionized, high- or low-excitation, atomic or molecular) to varying degrees and their effects should be possible to identify and calibrate also in the spectrum of the full stellar disk. 

Such an approach is made feasible by two developments.  Firstly, on the theoretical side, convectively driven gas motions near stellar surfaces can now be modeled in detail with time-dependent 3D [magneto-]hydrodynamics, and such simulations are now established as quite realistic descriptions of the spectral-line forming regions in the photospheres of at least solar-type stars.  A more recent development is the computation of complete stellar spectra from such models; \citet{chiavassaetal18, dravinsetal21a, dravinsetal21b}; Ludwig et al.\ (in prep.).  These produce complete and complex stellar spectra, incorporating a huge number of spectral lines from databases such as VALD\footnote{VALD, The Vienna Atomic Line Database of atomic and molecular transition parameters of astronomical interest} and others \citep{heiteretal15, ryabchikovaetal15}, and are able to so do so in sampling wavelengths with a hyper-high spectral resolution ($\lambda$/$\Delta\lambda$\,$\sim$1,000,000)\footnote{The term hyper-high is used to denote spectral resolutions $\lambda$/$\Delta\lambda\gtrsim$\,10$^6$ since ultra-high is already in common use for describing spectrometers with the much lower resolutions of `only' $\sim$200,000.}, spanning all observable spectral ranges. Details in the resulting spectra depend on a complex interplay between, for example, atmospheric opacity in different atmospheric inhomogeneities, blending of multiple spectral lines, and as seen under different viewing angles across the stellar disk in different wavelength regions.  Of course, the modeling entails various approximations, but since the synthetic spectra are free from random photometric noise, spectral microvariability such as the radial-velocity jittering of different spectral features can be followed throughout the simulation sequences by measuring the relative wavelength displacements of any feature anywhere in the spectrum and correlations with other spectral parameters identified.

\subsection  {Photometric irradiance flickering}  

Corresponding challenges in microvariability exist for the photometric flickering, a limiting parameter in identifying photometric transits of small planets across stellar disks and also, more generally, a diagnostic tool for studying the presence of granulation on stellar surfaces, as predicted from hydrodynamic models for different classes of stars  \citep[e.g.][]{ludwig06, lundkvistetal21}.  With $\sim$10$^6$ granules across the solar disk, typical brightness contrasts of $\sim$20\,\% in each granule reduces to $\sim$2\,10$^{-4}$ after averaging over such a million random locations \citep[e.g.][]{nemecetal20}.  Precision photometry from space reveals the temporal power spectra of such flickering, and can be used to infer stellar properties \citep{cranmeretal14, kallingeretal14, samadietal13a, samadietal13b, sulisetal20}.

\begin{table*}
\caption{Identifiers for the CO\,$^5$BOLD models used to compute synthetic solar spectra.  The identifiers are unique and are also used in other work.  The spectra were sampled in 47\,$\times$\,47 grids across the simulation areas of 140\,$\times$\,140 horizontal spatial points extending over $hsize$ in $x$ and $y$, at 19 or 20 temporal snapshots, and for up to 21 angular directions.}             
\label{table:co5bold_models}      
\centering          
\begin{tabular}{c c c c c c c}
\hline\hline      
\noalign{\smallskip} 
T$_{\textrm{eff}}$ [K] & log~$\varg$ [cgs] & Temp.\ RMS [K] \ & Model & Spectral points & boxsize:$hsize$ [cm] & boxsize:$vsize$ [cm] \\ 
\noalign{\smallskip}
\hline  
\noalign{\smallskip}
5782 & 4.44 & 12  & N58 - d3gt57g44n58  & 47 $\times$  47  $\times$ 19 $\times$ 13 & 5.60 10$^{8}$ &  2.25 10$^{8}$ \\
5774 & 4.44 & 16  & N59 - d3gt57g44n59  & 47 $\times$  47  $\times$ 20 $\times$ 21 & 5.60 10$^{8}$ &  2.25 10$^{8}$ \\
\noalign{\smallskip}
\hline                  
\end{tabular}
\end{table*}

\begin{figure}
 \centering
 \includegraphics[width=\hsize]{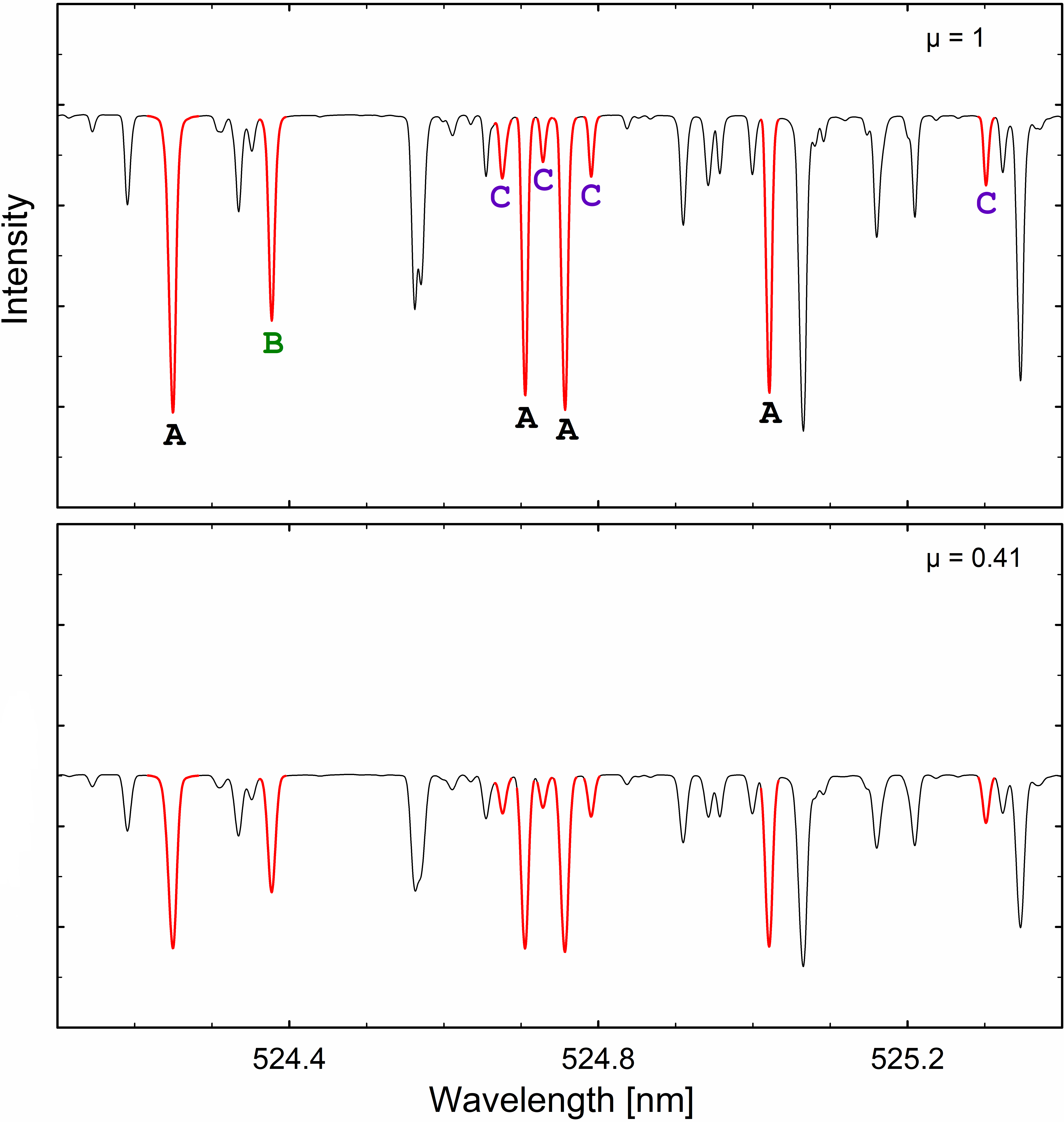}
     \caption{Examples of selected lines, here some \ion{Fe}{i} lines around $\lambda$~525 nm from the N59 model.  Top: synthetic spectra at solar disk center, $\mu$ = 1, bottom: $\mu$ = 0.41.  Clean lines selected for detailed analysis are marked in red (strong-medium-weak marked as A-B-C).  Lines become noticeably broader toward the limb.}  
\label{fig:co5bold_525nm}
\end{figure}

\begin{figure*}
\sidecaption
 \centering
 \includegraphics[width=18cm]{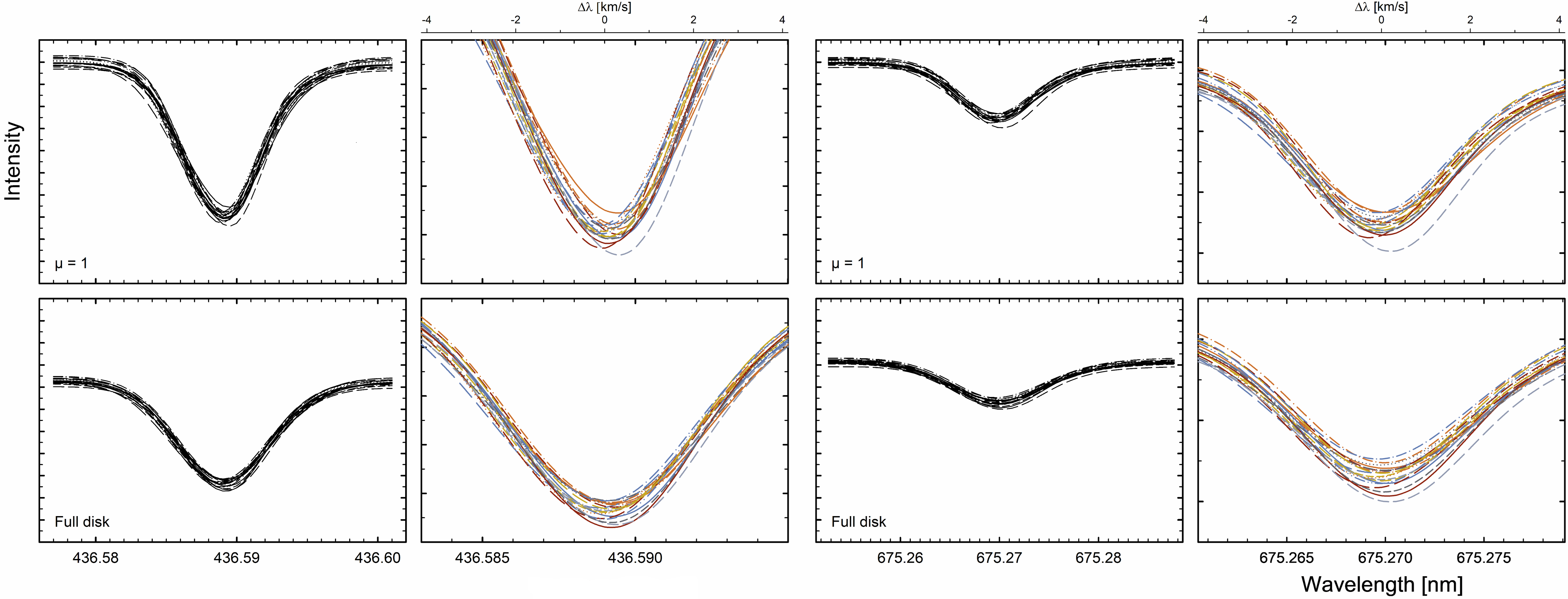}
    \caption{Example of temporal variability of line profiles during the N59 simulation sequence, sampled at 20 instances of time.  Left columns: the strong \ion{Fe}{i} $\lambda$\,436.59 nm, $\chi$ = 2.99 eV line in the blue; Right columns: the weak \ion{Fe}{i} $\lambda$\,675.27 nm, $\chi$ = 4.64 eV line in the red.  Upper row shows profiles for disk center $\mu$ = 1; bottom row shows integrated sunlight, corresponding to a full solar disk, synthesized using center-to-limb profiles from the small simulation area.  Zoomed-up frames to the right show line centers on a uniform `wavelength' scale in units of radial velocity (at top).  In the zoomed-up frames, tick marks on the vertical axes correspond to those for the full profiles.  }  
    \label{fig:resolved_profiles_n59}
\end{figure*}

\subsection{Limitations of theory}

Ideally, one could envision detailed magnetohydrodynamic simulations embracing entire stellar surfaces, producing time-variable spectral atlases for varying mixtures of different types of surface structures on different classes of stars.  This is not yet practical and the scope has to be limited.  We start with the solar spectrum, both because it is the most studied and modeled one, and because theoretical predictions can be observationally tested in quite some detail.  Realistic 3D simulation volumes can cover groups of multiple photospheric granules and permit to follow how their appearance changes when seen under different angles, from disk center out toward the limb.  Still, simulated solar surface areas can cover only a small fraction of the full surface.  

As described in Sect.\,5 below, the equivalent of a full solar disk is obtained from this simulation area of 5.6\,$\times$\,5.6 Mm$^2$ by summing appropriately weighted spectra from various computed center-to-limb angles, producing a spectrum that retains the fluctuations from the small simulation area.  The actual solar hemisphere area is larger by a factor of $\sim$100,000 and the projected solar disk by $\sim$50,000.  In case of statistically independent random fluctuations (neglecting details of geometric projection effects and statistical weighting by limb darkening), the averaged full-disk amplitudes should then decrease by a factor equal to the square root of this number.  However, away from disk center, spectra are also computed for up to eight different azimuthal directions, thus providing some further averaging.  Nevertheless, even if such spectra are dissimilar as seen from various oblique directions, such samples are from the same instance in time and cannot be seen as statistically fully independent.  While it is awkward to precisely quantify their level of randomness, the amplitudes for the actual full solar disk can be expected to average out to $\sim$150-200 times smaller than those from our modest simulation volume.  As will be seen below, the jittering of spectral lines averaged over the simulation area typically extend over $\sim$300 m\,s$^{-1}$, which would then reduce to $\sim$1-2 m\,s$^{-1}$ in actual integrated sunlight.  Such are the magnitudes of microvariability to be studied here. 

The modeling discussed in this first paper concerns mainly the non-magnetic granulation.  This covers most of the solar surface, and provides most of the solar flux.  Analyses of the time-variable spectra from the simulation volume enable to identify relations between various parameters and to extrapolate what amounts should remain in actual full-disk spectra.  However, to properly reproduce also larger-scale features such as supergranular flows or p-mode oscillations will require (much) larger simulation volumes.  Also, a significant part of the spectrum variability originates in regions of magnetic granulation and spots.  Magnetohydrodynamic 3D simulations for such are available and their synthetic spectra can be computed, foreseen to be discussed in future papers.  

Irrespective of the computational sophistication, the accuracy of synthetic spectra is ultimately limited by the quality of atomic and molecular data.  Current synthetic spectra include more than half a million spectral lines, with their wavelengths and oscillator strengths originating from laboratory measurements or theoretical predictions.  The accuracy of those data, originating from a mixed variety of sources, is somewhat variable.  While the overall spectrum does reproduce the observed one to a rather high degree, detailed examination often reveals certain differences in relative line strengths, of blending lines, or other.  For some studies, these limitations may be circumvented: e.g., to gain a knowledge of the absolute amount of convective blueshifts, an identical spectral synthesis may be repeated with the same input data but for a static model atmosphere, where convective lineshifts cannot exist, and the difference to the hydrodynamic case then evaluated \citep{dravinsetal21a}.  In a somewhat analogous vein, we do not need to precisely reproduce the exact strength of each individual spectral feature but we rather examine the temporal jittering in wavelength and changes in shapes and strengths of representative types of lines, even if their absolute wavelengths or exact oscillator strengths would not be precisely known.

\begin{figure*}
\sidecaption
 \centering
 \includegraphics[width=18cm]{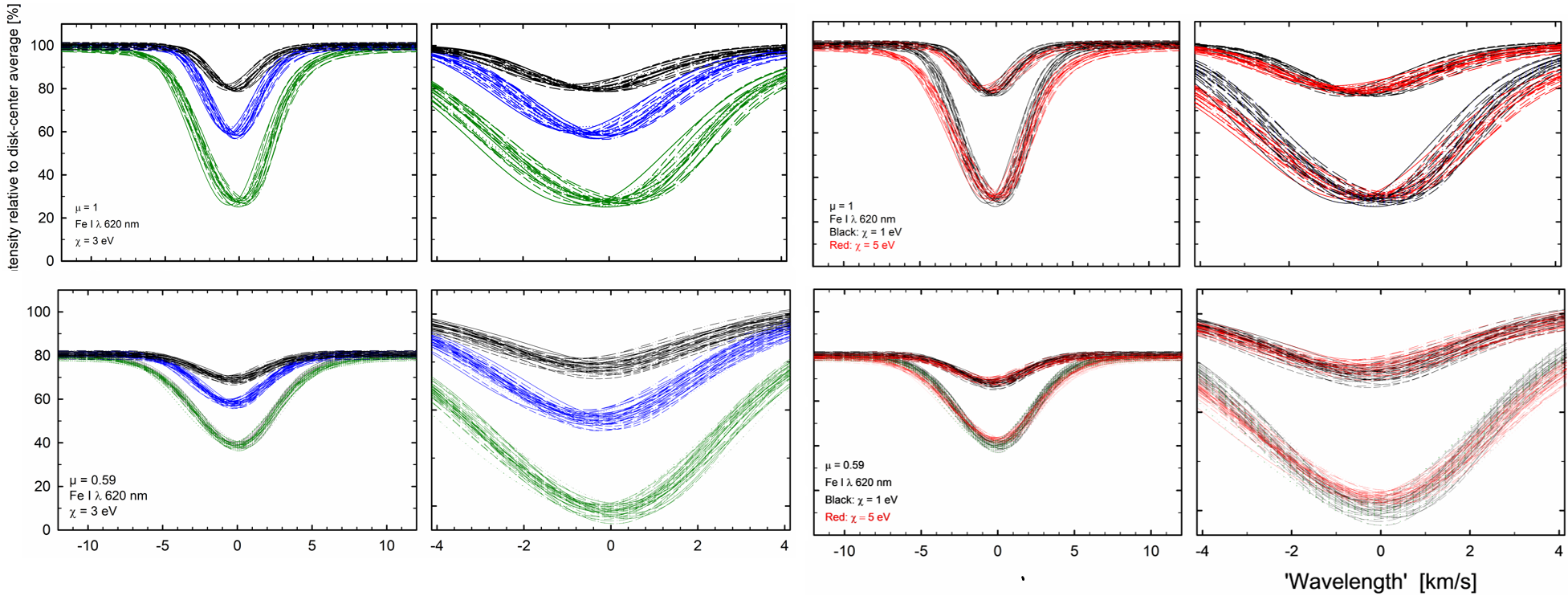}
    \caption{Temporal variability of idealized line profiles during the N58 simulation sequence, sampled at 19 instances of time.  Left columns: Differently strong \ion{Fe}{i} lines at $\lambda$\,620 nm, $\chi$ = 3 eV; Right columns: Lines of similar strength but with different lower excitation potentials,  $\chi$ = 1 eV (black) and 5 eV (red).  Top row: Profiles at disk center $\mu$ = 1; bottom row: $\mu$ = 0.59.  To limit cluttering, profiles are shown for only two (orthogonal) azimuth angles.  The right-hand frames show the line centers zoomed up, with `wavelength' scales equal to those in the previous Figure \ref{fig:resolved_profiles_n59}.  In these zoomed-up frames, tick marks on the vertical axes correspond to those for the full profiles.}  
    \label{fig:resolved_profiles_n58}
\end{figure*}

\begin{figure*}
 \sidecaption
 \includegraphics[width=11cm]{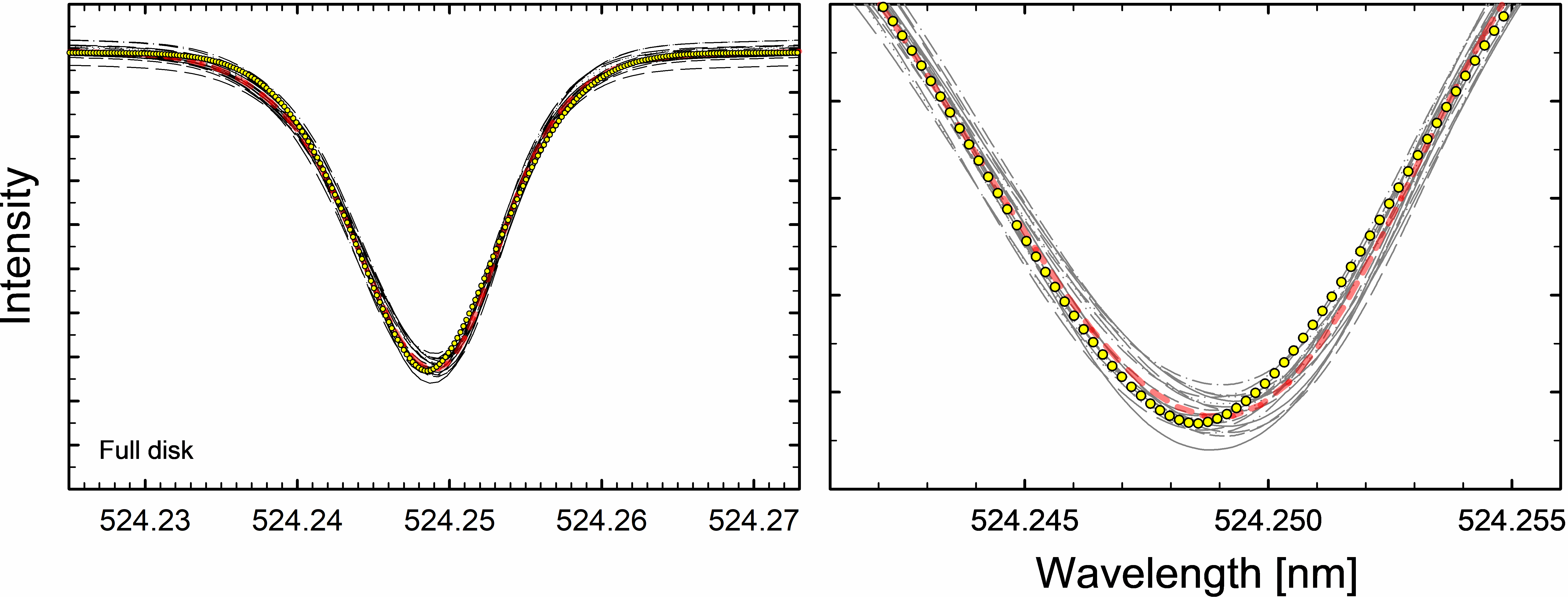}
     \caption{Example of 5-parameter Gaussian-type fit to a typical \ion{Fe}{i} line profile in integrated sunlight.  Profiles from each temporal snapshot are thin black lines; their average is bold-dashed red, and the fitted profile is yellow circles.  In the zoomed-up frame at right, tick marks of the vertical scale are carried over from the left frame. }  
\label{fig:gaussianfit}
\end{figure*}

\subsection{Realistic project aims}

If microvariability signatures generated in stellar atmospheres can be adequately understood, one could then determine what observational efforts eventually would be required to detect and calibrate those.  Not less motivating for both the theoretical and observational aspects, is that both of them are expected to improve in the near future.  With enhanced computational capacity, numerical simulations are being extended to greater solar and stellar volumes, also incorporating an increased variety of magnetic features, sunspot umbrae, penumbrae, and other.  Observationally, several light feeds from solar telescopes are now connected to extreme-precision radial velocity spectrometers and are expected to continue collecting data over at least a solar activity cycle or longer.  Of additional interest will be data from across a spatially somewhat resolved solar disk measured with comparable precision, e.g., monitoring the characteristics in wavelength jittering in quiet areas versus magnetic ones.  Such behavior can be theoretically predicted but corresponding observations are not yet available, although projects are in progress \citep{leiteetal22}.

The variability of the optical spectrum of the Sun seen as a star has previously been studied mainly in features such as \ion{Ca}{ii}~H\,\&\,K or H$\alpha$ but signals in such chromospheric lines are unlikely to precisely relate to radial velocities, where the signal comes from photospheric lines only.  However, the number of photospheric features that potentially can be examined, combined or correlated with one another is enormous and here only a limited number of selected samples can be surveyed.  The aim will be to outline possible paths toward the identification of spectral signatures to serve as proxies for calibrating (at least part of) the radial-velocity jittering in sunlight, and eventually also for other stars.  Such proxies would need to be measurable at ground-based observatories independently of the overall radial velocity.  These might include, e.g., velocity differences between various line-groups, changes in line-depths or widths, or perhaps even broad-band photometric colors.  A typical photospheric absorption line subtends a wavelength width of $\sim$10~km\,s$^{-1}$, and a radial-velocity signal of 1~m\,s$^{-1}$ thus amounts to only 1 part in 10,000.  Quite possibly, proxies for such signals are comparably minute and probably may require extensive modeling to be securely identified. 

The value of understanding photospheric spectrum microvariability extends to also areas outside monitoring stellar radial motion, such as searches for varying fundamental constants in nature, using precise differences of spectral-line wavelengths in solar-type stars \citep{berkeetal23a,berkeetal23b, murphyetal22}.

\begin{figure*}
\sidecaption
 \includegraphics[width=11cm]{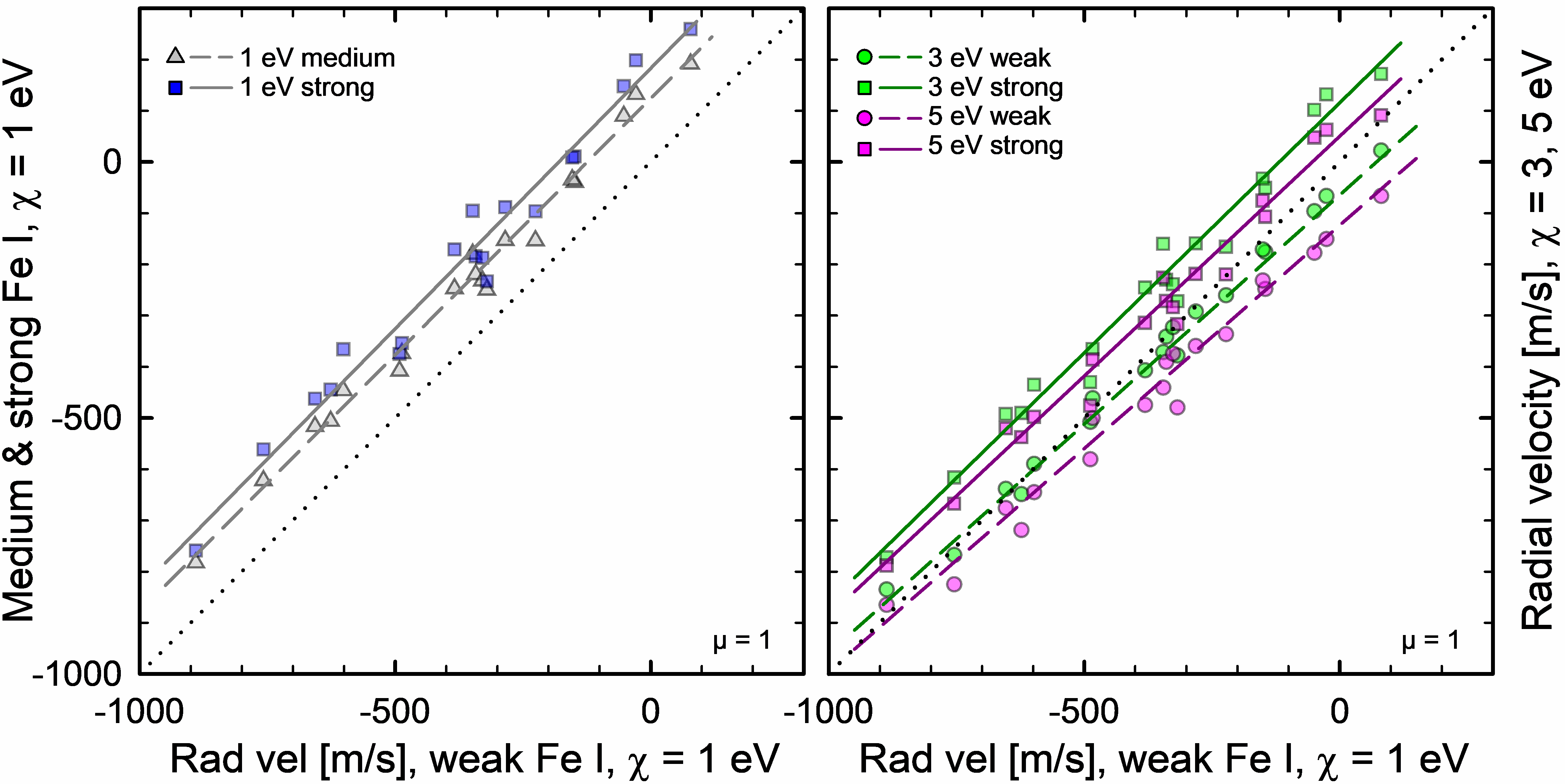}
     \caption{Radial-velocity shifts of synthetic isolated \ion{Fe}{i} lines at $\lambda$\,620 nm in the N58 model.  Relations between shifts at solar disk center, $\mu$ = 1, for weak, medium and strong lines, are shown for different excitation potentials.  Each point represents one temporal snapshot during the simulation.  Dotted reference lines mark the identity relations, the other fitted lines indicate how the dependence of convective blueshift on line-strength and excitation potential generally remains valid also when the spatially averaged radial velocity fluctuates. }  
\label{fig:radvel_chi_a}
\end{figure*}

\section{CO\,$^5$BOLD modeling of the solar spectrum}

Building upon previous work \citep{dravinsetal17a, dravinsetal17b, dravinsetal18, dravinsetal21a, dravinsetal21b}, profiles of synthetic spectral lines are computed from 3D atmospheric structures produced in hydrodynamic simulations, using those as grids of space- and time-varying model atmospheres.  Two different non-magnetic solar models are used here: N58 and N59 (Table \ref{table:co5bold_models}).  Both these have solar temperature, surface gravity and metallicity but differ in some computational aspects, and the details of their granular evolution geometries are somewhat different (although, of course, statistically similar). Computationally, the N59 model is a successor to N58, which was examined in \citet{dravinsetal21a, dravinsetal21b}.  For example, while the radiation balance in N58 was calculated using five wavelength bins, N59 uses 12 and hence, the photospheric energy budget should be better represented.   For each model, the horizontal simulation area is 5.6\,$\times$\,5.6 Mm$^{2}$, with each modeling pixel in the 140\,$\times$\,140 grid of hydrodynamic simulation equal to 40\,$\times$\,40\,km$^{2}$ (Table~1).  In the simulation sequences, the gradual evolution of granular structures can be followed.  Their duration of about two solar hours permitted some 20 temporal snapshots to be selected sufficiently far apart in time to ensure largely uncorrelated granulation flow patterns between successive samples.  For such snapshots, spectra for all different elevation and azimuth angles are computed on a grid of 47\,$\times$\,47 points (each horizontally 120\,$\times$\,120\,km$^{2}$), downsampled by merging 3\,$\times$\,3 pixels in the hydrodynamic simulation.  Since successive snapshots are chosen well apart in time, their later averaging should have a largely random character but, for a given snapshot, the spectra at different emerging angles originate from the same granular structures (viewed from different directions), making their level of mutual randomness more awkward to estimate. 

Details about the computation of these types of models are in \citet{tremblayetal13}, where also representative synthetic granulation images are shown, both for models at solar temperature, and others.  Further details are in Ludwig et al.\ (in prep).  

From the N58 model, profiles of single, idealized and isolated spectral lines for \ion{Fe}{i} at $\lambda$\,620 nm were computed for five different line strengths and three different lower excitation potentials, $\chi$ = 1, 3, and 5 eV.  Four different center-to-limb positions $\mu$ = cos\,$\theta$ = 1.00, 0.87, 0.59, 0.21, and four azimuth angles $\psi$ = 0, $90^\circ$, $180^\circ$ and $270^\circ$ were treated in spectral line synthesis carried out in local thermodynamic equilibrium (LTE).  Such calculations were made for 19 temporally sampled snapshots during the extent of the simulation.  Each profile was computed for 201 wavelength points covering $\pm$15~km\,s$^{-1}$.  The sampling stepsize thus corresponds to a hyper-high spectral resolution $\lambda$/$\Delta\lambda$ = 2,000,000.  Hyper-high resolution is required to fully resolve intrinsic line asymmetries and to resolve wavelength shifts on levels of cm\,s$^{-1}$.  Spectra from this series have been discussed in \citet{dravinsetal17a, dravinsetal17b, dravinsetal18}, and also in \citet{dravinsetal21b}.  

For the N59 model, complete and complex spectra were computed, incorporating practically all relevant spectral lines from databases such as VALD.  Calculations were made for the spectral range of 120-1000 nm, and for 20 largely uncorrelated instances in time during the simulation.  The wavelength sampling corresponds to the hyper-high value $\lambda$/$\Delta\lambda$ $\sim$900,000. For each instance in time, four center-to-limb angles $\mu$ = 1.0, 0.79, 0.41, and 0.09 are provided.  The spectra were computed for a total of 21 emergent directions: eight azimuth angles in steps of $45^\circ$  for $\mu$ = 0.79 and 0.41, and four angles in steps of $90^\circ$ for $\mu$ = 0.09.  With intensities in absolute units, it becomes possible to obtain also synthetic broad-band colors.  Time-averaged spectra from analogous types of calculations were presented previously in \citet{dravinsetal21a, dravinsetal21b} for different classes of stars, but now the spectra are time-resolved. 

Despite the spatial averaging over the model area, the total volume of these data is significant: with some 3 million data points in each spectral realization, and for each of the 21 position angles, at each of the 20 instances in time, the total amounts to more than a billion spectral intensity values, requiring some selectivity in the handling of spectra and in choosing and limiting lines for analysis.  In the figures, plots for specific center-to-limb positions refer to averages over all computed azimuth angles, unless noted otherwise.  Integrated sunlight is computed as for a full disk (and labeled thus in the figures) by appropriately interpolating and summing spectra from different $\mu$-values (Fig.\ref{fig:full_spectrum}).  The numerical values, however, refer to those obtained from the small simulation area while -- as discussed above -- those for the actual Sun are  estimated to be smaller by a factor $\sim$150-200.   Further degrees of freedom are offered by stellar rotational broadening; such variants were treated in \citet{dravinsetal17a, dravinsetal17b} but here we limit ourselves to rotationally unbroadened lines only.   

Each of these N58 and N59 model spectra offers its advantages and limitations.  The advantage with idealized single lines in N58 is that these can highlight possible subtle contrasts between lines of only slightly differing strengths or excitation potentials.  However, such lines might not represent observational realities, where not all combinations of line parameters are present and with often multiple blending lines in crowded spectra.  Such complete spectra are instead produced in N59 but require some effort in selecting representative and distinct spectral features.  With more than 600,000 lines included, any selected line will unavoidably incorporate also some weak blends which might have a different temporal behavior than the main line, thus adding some astrophysical noise.  The complete spectra in N59 should illustrate what could be practically measurable while the idealized lines in N58 should help to understand the physical origins of possibly identified relations. 

A limitation comes from the circumstance that reliable modeling is for the lower photospheric levels in the atmosphere while higher chromospheric structures are not treated.  Consequently, lines that are (partially) formed in the uppermost photosphere or lower chromosphere cannot be expected to be precisely reproduced.  On one hand, such chromospherically influenced lines -- perhaps \ion{Na}{i} D, the \ion{Mg}{i} triplet or Balmer lines from hydrogen -- are not likely to be used for measuring precision radial velocities.  However, we will still examine their behavior since their modeling may confirm trends when going to stronger lines formed in higher atmospheric layers and they should follow photospheric events more closely than chromospheric \ion{Ca}{ii}~H\,{\&}\,K-line emission.

Table\ref{table:co5bold_models} gives the unique identifiers of the models, which enables comparisons with other work based on the CO\,$^5$BOLD grid \citep{freytagetal12}. The number of sampling points for each spectral line profile is the product of horizontal spatial resolution points in x and y, the number of temporal snapshots extracted during the simulation sequence, and the number of different directions (combined elevation and azimuth angles) for the emerging radiation. The x-y horizontal {\it{hsize}} and z vertical  {\it{vsize}} extents of the hydrodynamic simulation volume are there given in centimeters.

\begin{figure*}
\sidecaption
 \includegraphics[width=11cm]{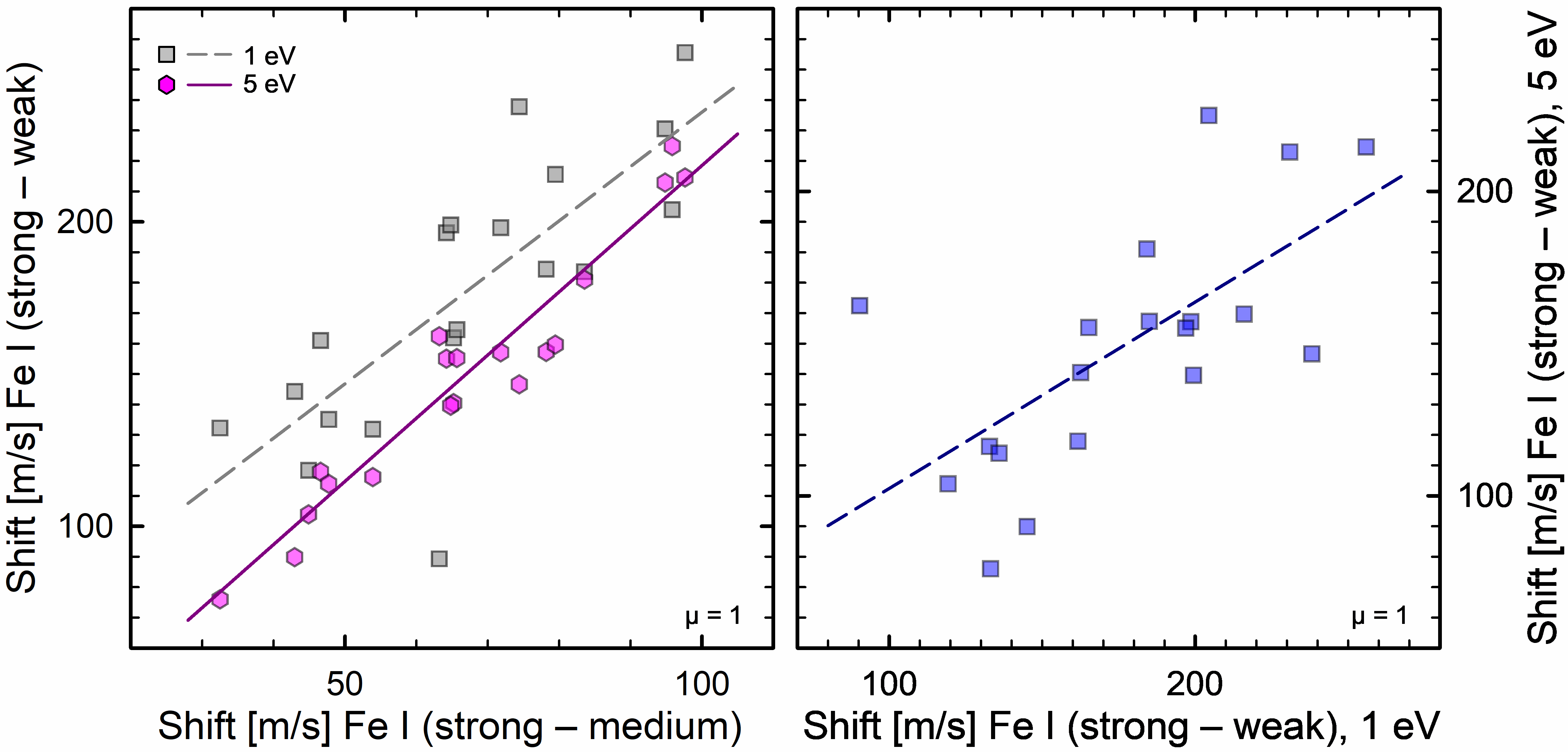}
     \caption{Differences in radial-velocity shifts between lines with different strengths (but similar excitation potentials).  These data at $\mu$ = 1 are from synthetic isolated \ion{Fe}{i} lines at $\lambda$\,620 nm in the N58 model.  Each point represents the difference between instantaneous radial velocities (e.g., strong line minus weak) at one temporal snapshot during the simulation. Straight lines are fits to the data points. Different scales between the frames are adapted to reflect how convective blueshifts deviate for lines with increasingly different strengths. }  
\label{fig:radvel_chi-difference}
\end{figure*}

\section{Synthetic spectral lines}

Figure \ref{fig:full_spectrum} shows complete synthetic spectra for three center-to-limb angles from the N59 model: disk center $\mu$\,=\,1, $\mu$\,=\,0.41, and  $\mu$\,=\,0.09.  These data are from one particular instance in time; however the differences between successive temporal snapshots or between different azimuth angles are much too small to be visible on this scale.  

Figure \ref{fig:co5bold_525nm} gives an example of the selection of differently strong \ion{Fe}{i} lines in one particular wavelength region while Figures \ref{fig:resolved_profiles_n59} and \ref{fig:resolved_profiles_n58} illustrate the temporal variability in lines from the N59 and N58 models.

\subsection{Fitting line parameters}

To enable searches for dependences among various line parameters, the information resident in the line profiles must somehow be parameterized. We stress that the line profiles from each instant in time are not the spatially resolved ones.  Such local profiles, originating from specific granular inhomogeneities, display grossly varying shapes \citep[e.g.,][]{asplund05, bergemannetal19, dravinsnordlund90}, and (even disregarding the data quantity) would not lend themselves to easily understandable quantification.  Each of the current temporally resolved line profiles was spatially averaged over more than 2000 spatial points across the hydrodynamically modeled stellar surface, embracing numerous granulation structures, thereafter resulting in profiles that are closely reminiscent of the temporally averaged ones, and that can be described with commonly used parameters.  However, these still retain noticeable variability over time since the simulation area is tiny compared to the full solar disk. 

To quantify these profiles, various tests of line-fitting parameters were made, finding a suitable approximation by fitting with five-parameter Gaussian-type functions of the type $y_0 - a\cdot\exp[-0.5\cdot(|x-x_0|/W)^c]$, yielding $y_0$ as the continuum level; $a$ for the line's absorption depth; $W$ as a measure of the line width, $c$ indicating the degree of the line’s boxiness or pointedness, and $x_0$ for the central wavelength.

Although the profiles of course are not precisely Gaussian, these functions appear reliable to use.  Different varieties were tested (Lorentzian, etc.), but it was found that such Gaussian-type functions gave robust estimates for all basic line parameters.  For the wavelength positions, this fitting thus provides one value that refers to the displacement of the line as a whole.  For asymmetric profiles, a single number for the line wavelength is not a uniquely defined measure, but depends on whether the fit is made to the central or more extended parts of the profile, and with what type of weighting.  However, the physical variability among our line profiles is much greater than possible ambiguities stemming from the exact shape of the fitting functions, and -- in any case -- here we do not aim at determining absolute line quantities as in \citet{dravinsetal21a}, but only their relative changes in time.  Fig.\ref{fig:gaussianfit} gives a representative example of such a fit.

With a significantly greater sample (producing more stable averages), one could parameterize with respect to also line asymmetries or correlations between different portions of the line profile \citep[e.g.,][]{ludwigsteffen08}.  However, studies of more subtle effects are currently constrained by the limited number of temporal samples.  Data from ab-initio simulations, where the character of the output cannot be precisely predicted, have a character similar to that of noisy observational data in that they need to be arranged and averaged in order to discern trends.  The parameters of wavelength shift and line width appear to be the most readily measurable quantities (also observationally), likewise for the center-to-limb dependences in spatially resolved spectra.

\begin{figure*}
\sidecaption
 \includegraphics[width=11cm]{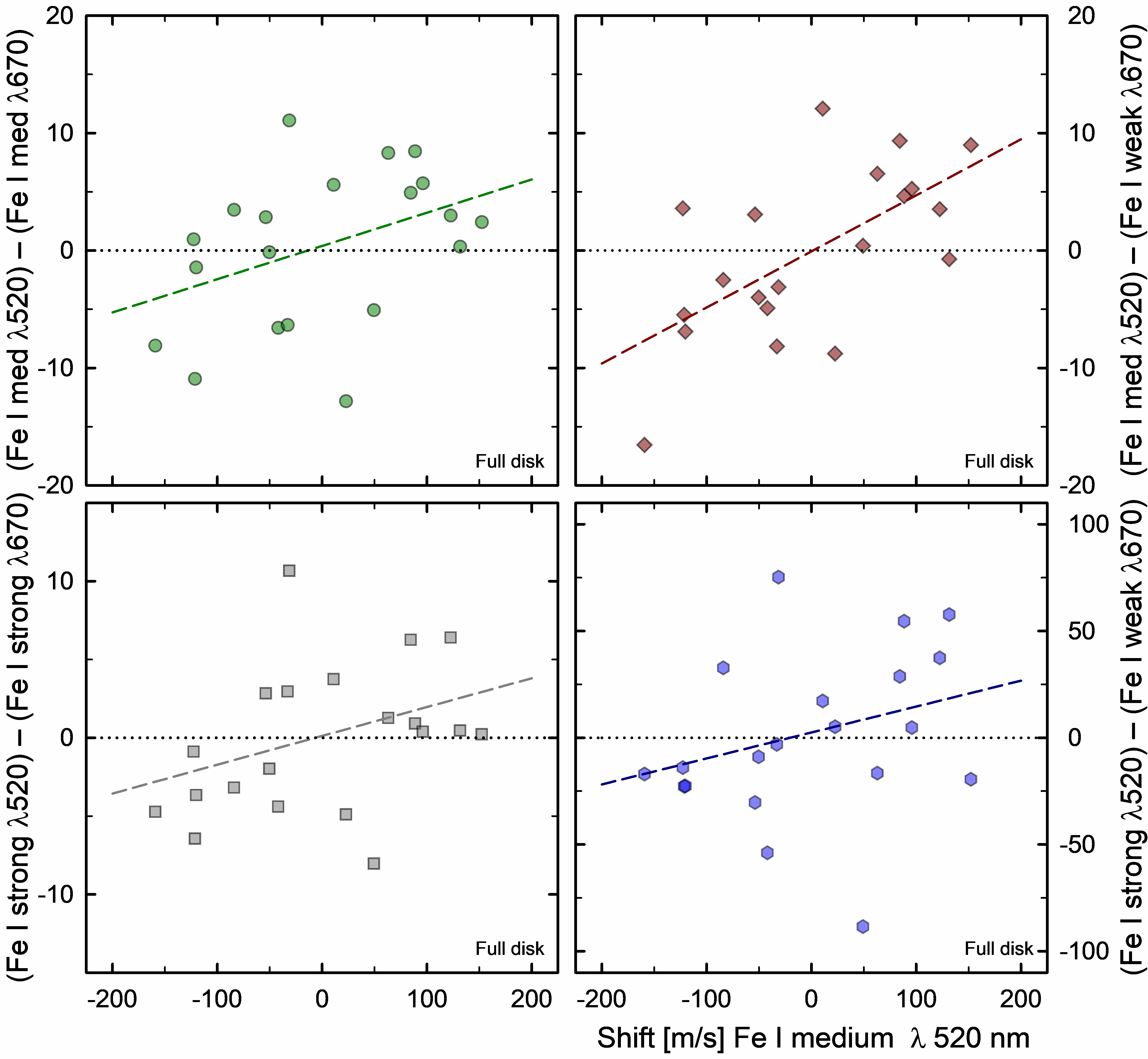}
     \caption{Differences in radial-velocity shifts between \ion{Fe}{i} lines of different strength in different spectral regions of integrated sunlight, versus shifts in medium-strong \ion{Fe}{i} lines at $\lambda$\,520 nm.  Each point represents one temporal snapshot during the N59 simulation.  The spectra correspond to those of a full solar disk, as synthesized from a sequence of center-to-limb spectra, although the numerical values [m\,s$^{-1}$] refer to the small simulation area (for the actual Sun, divide by  a factor $\sim$150-200).  Note that the vertical scales differ among the frames.  The fitted straight lines suggest the trends but also indicate the limitation from finite simulation sequences.  }  
\label{fig:diff_shifts_520nm}
\end{figure*}

\begin{figure*}
\sidecaption
 \includegraphics[width=11cm]{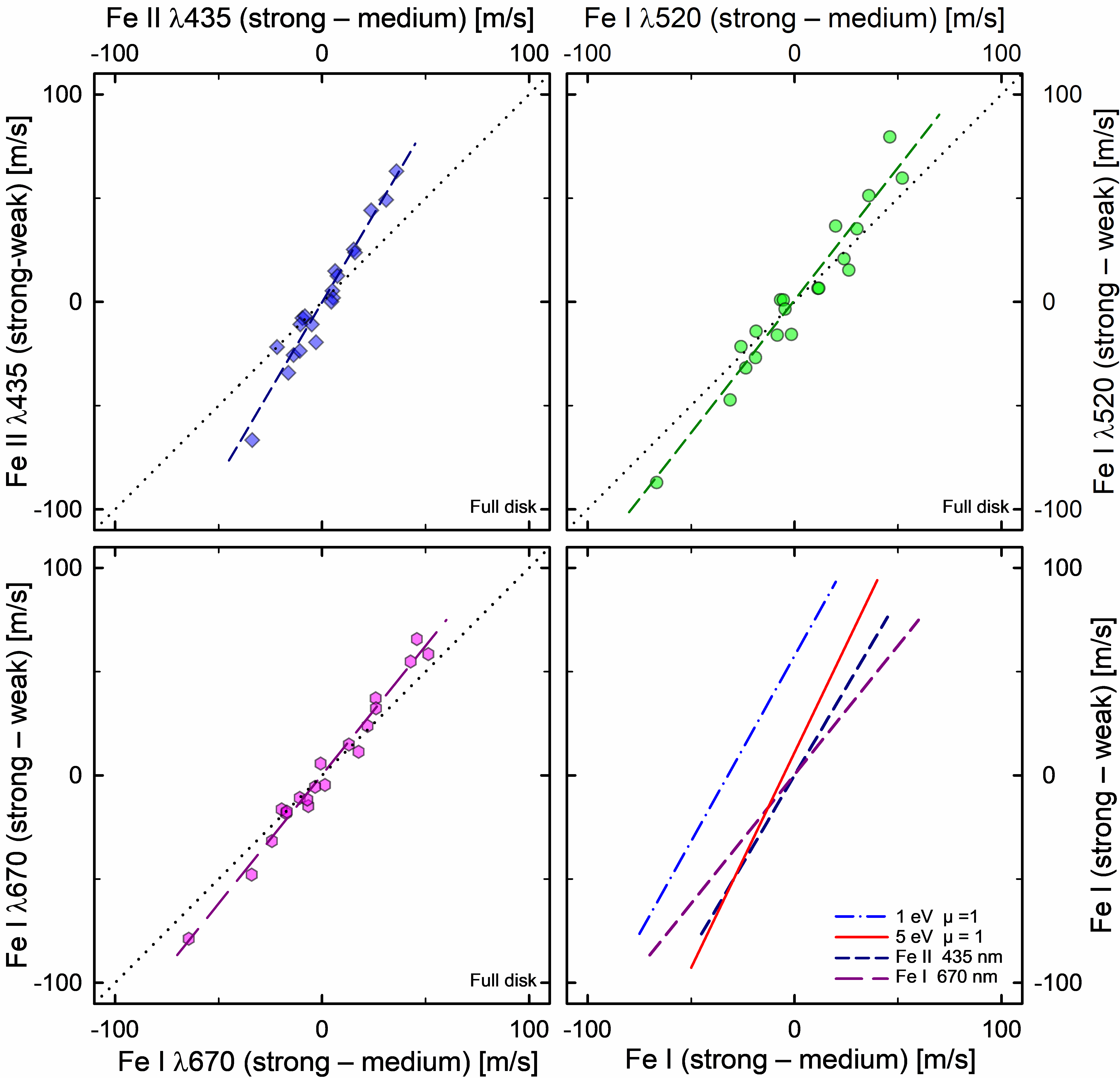}
     \caption{Radial-velocity shifts between \ion{Fe}{i} and \ion{Fe}{ii} lines of relatively similar strengths (strong minus medium) versus those with greater differences (strong minus weak), in different spectral regions of integrated sunlight.   The numerical values refer to the small simulation area in the N59 model (for the actual Sun, divide by  a factor $\sim$150-200).  All frames have identical scales. }  
\label{fig:diff-diff}
\end{figure*}

\subsection{Variability in radial velocity and irradiance}

Considerable efforts have been made in studying the Sun as a star and in searching for statistical relations between various parameters of solar and stellar variability, including work by \citet{ceglaetal19}, \citet{colliercameronetal19, colliercameronetal21}, \citet{debeursetal22}, \citet{dumusqueetal15, dumusqueetal21}, \citet{haywoodetal16}, \citet{lanzaetal16, lanzaetal19}, \citet{maldonadoetal19}, \citet{marchwinskietal15}, \citet{thompsonetal20}; references therein, and also as reviewed by \citet{haraford23}.  These studies use various statistical approaches to search for possible correlations and empirical proxies for radial-velocity fluctuations.  Considerations of other spectral types include \citet{dumusqueetal11a} and \citet{meunieretal17a, meunieretal17b}. 

An estimate of the trends in our models indicates that excursions around $\pm$100 m\,s$^{-1}$ correspond to about $\pm$1\% in irradiance.  As discussed above, actual full-disk amplitudes will be reduced from the simulation area by a factor of $\sim$\,200 or 150, leading to about 0.5 or 0.7 m\,s$^{-1}$ per 50 ppm in irradiance.  These values seem consistent with those deduced by \citet{ceglaetal19}, \citet{meunieretal15}, and \citet{meunierlagrange20} for the granulation-induced signal in apparent solar full-disk radial velocity although further contributions may come from larger-scale structures such as supergranulation \citep{meunierlagrange19}, not yet accessible to detailed 3D modeling.  A further modeling caution is that brightness variations are also driven by p-mode oscillations which are incompletely treated in the small simulation volumes.

We also note recent related work by \citet{Kitiashvilietal23}, where synthetic radial-velocity fluctuations are computed from a 3D model of the G0~V star HD209458, revealing amplitudes of $\sim\pm$200~m\,s$^{-1}$ over their simulation area of 12.8 Mm width.

\section{Photospheric \ion{Fe}{i}  and \ion{Fe}{ii}  lines}

An examination of the simulation sequences shows that different absorption lines generally fluctuate in phase although with subtle (but systematic) differences between lines of different strength or located in different spectral regions.  The absolute amounts of convective blueshift in time-averaged spectra of course show their familiar dependence on line-strength \citep[e.g.,][]{dravinsetal21a} but there are also line-strength dependences of the amplitudes in the temporal variation of these blueshifts. 

To identify systematics in the spectral-line jittering, various correlations were examined between different groups of lines, using the parameters determined from the five-parameter Gaussian-type fitting as described above.  From the N58 model, idealized single \ion{Fe}{i} lines at $\lambda$\,620 nm were computed with a nominal zero rest wavelength shift, and thus the instantaneous radial velocity value equals the displacement relative to this reference point of zero.  Five different line-strengths were computed in N58, each for the three different lower excitation potentials of 1, 3, and 5 eV.  In all cases, the trend of fitted parameters with changing line-strengths or excitation potentials was found to be monotonic and in the following figures, only a few representative cases are selected to be plotted.

From the N59 model with complete spectra, groups of clean \ion{Fe}{i} lines were selected in representative wavelength regions: around $\lambda$\,430 nm, $\lambda$\,520 nm (Fig.\ \ref{fig:co5bold_525nm}), and $\lambda$\,670 nm, and an \ion{Fe}{ii} sample around $\lambda$\,435 nm.  These regions were selected to span different wavelength regions, where also lines of different relative strength were present (labeled as strong-medium-weak).  However, the absolute line strengths differ among the regions: \ion{Fe}{i} lines in the red are rather weaker than lines in the blue (Fig.\ \ref{fig:resolved_profiles_n59}), and at longer wavelengths there are too few \ion{Fe}{ii} lines for a meaningful sample.  Also, a further subdivision in N59 with respect to excitation potential was not seen as practical.  The instantaneous radial velocity value for each of those lines is obtained as the displacement relative to each line's average over the simulation sequence. 

The number of temporally separated snapshots during the simulation sequences is 19 for the N58 model, and 20 for N59.  Each snapshot represents the spatial average over the simulation area, and these are the number of points plotted in the figures below.  The values shown are averages for the selected lines from each region; however it was found that, within each wavelength region, the radial-velocity shifts are almost identical for different lines with the same strength; differences appear only between different line-strengths, different wavelength regions, and different excitation potentials.  This also implies that the noise in these lineshift plots originates from temporal fluctuations in the granulation structure during the 3D-simulation, not photometric or other uncertainties in the computed spectra.  To decrease such noise requires spatially and/or temporally more extended simulations.

\begin{figure*}
 \sidecaption
 \includegraphics[width=12.9cm]{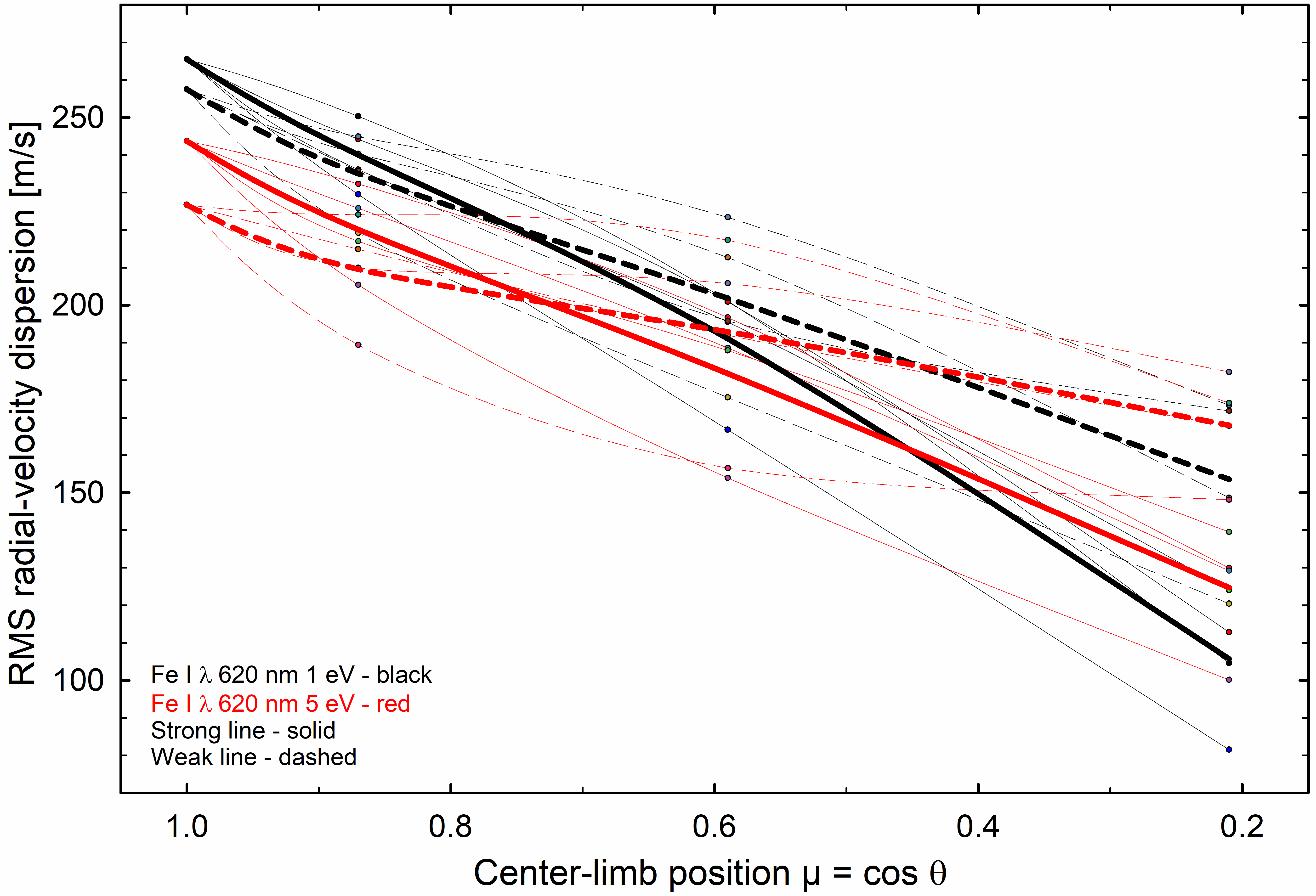}
     \caption{Spatially resolved jittering across the solar disk.  RMS values for radial velocities of idealized \ion{Fe}{i} lines at $\lambda$\,620 nm are shown for differently strong lines at excitation potentials $\chi$ = 1 and 5 eV.  Bold curves show averages over the small simulation area: solid for strong lines; dashed for weak.  Curves for $\chi$ = 1 eV are black, 5 eV are red.  Multiple thin lines illustrate the differences among different azimuth angles at each center-limb position. }  
\label{fig:idealized_lines_rms}
\end{figure*}

Figure \ref{fig:radvel_chi_a} shows fluctuations in the radial velocities as obtained from fits to isolated, differently strong \ion{Fe}{i} lines at $\lambda$\,620 nm in the N58 model; each point marks a different time during the simulation.  For reference, the horizontal axis shows the instantaneous radial velocity for a weak line with the excitation potential 1\,eV, while the plot is populated with data points for medium and strong lines, as in Fig.\ \ref{fig:resolved_profiles_n58}.  The fitted straight lines confirm that the convective blueshifts systematically decrease with increasing line-strength (left panel).  There is also a certain dependence on excitation potential (right panel), with higher-excitation lines exhibiting slightly greater blueshifts.  However, such trends seen for these idealized lines, are less obvious to identify in the real solar spectrum because not many really high-excitation lines of significant strength exist.  These plots are for solar disk center, $\mu$ = 1. 

Figure \ref{fig:radvel_chi-difference} compares differences in instantaneous radial-velocity shifts between pairs of idealized \ion{Fe}{i} lines at $\lambda$\,620 nm with modest differences in strength (strong minus medium; horizontal axis) versus simultaneous differences between more dissimilar lines (strong minus weak; vertical axis).  The plots show systematic trends, such that a given velocity shift between lines of rather similar strengths most often corresponds to a simultaneous but greater shift between lines with greater differences in strength.  There, amplitudes between weak and strong lines have some further dependence on excitation potential, although there is scatter among the limited number of data points.  

While Figs.\ \ref{fig:radvel_chi_a} and \ref{fig:radvel_chi-difference} are for idealized lines at solar disk center, Figure \ref{fig:diff_shifts_520nm} shows these types of relations for the simulated full solar disk, now measured for essentially unblended lines selected in the complete solar spectrum from the N59 model.   The spectrum for the integrated full solar disk is obtained by integrating over a disk, where the spectra from the four computed center-to-limb angles ($\mu$ = 1.0, 0.79, 0.41, 0.09) are first averaged over their respective azimuth angles, then interpolated to intermediate center-to-disk positions, and finally summed with appropriate wavelength-dependent limb-darkening values. This is equivalent to the spectrum of a tiny solar disk, where the fluctuation amplitudes of radial velocity, irradiance, and other still refer to the small simulation area.  As discussed above, the amplitudes for the actual full solar disk, will be $\sim$150-200 times smaller, although the exact number is awkward to precisely determine since it depends on understanding the degree to which temporal and azimuthal samples from the simulations can be considered statistically independent. 

As reference for the radial-velocity, the values for medium-strength lines at $\lambda$\,520 nm (Fig.\ \ref{fig:co5bold_525nm}) make up the horizontal axis in Fig.\ref{fig:diff_shifts_520nm}.   The vertical axes show simultaneous lineshift differences between pairs of line-groups with different strengths and/or in various wavelength regions.  Despite a scatter of points, all fitted lines indicate qualitatively similar trends, with the greatest amplitudes between strong lines around $\lambda$\,520 nm and weak ones in the far red at $\lambda$\,670 nm.

Figure \ref{fig:diff-diff} summarizes the main trends: velocity jittering in different lines is almost always in phase, but the amplitudes depend on line-strength, ionization level, and wavelength region.  The greatest amplitudes are noted for \ion{Fe}{ii}, while for \ion{Fe}{i}, both the jittering amplitudes and their differences between differently strong lines decrease for longer wavelengths in the red.  These data are from essentially unblended lines selected in the complete solar spectrum from the N59 model.  Straight lines at bottom right show the various fitted trends.  

As seen from the horizontal axis of Fig.\,\ref{fig:diff_shifts_520nm}, the total jittering amplitude over the simulation area is on the order of $\pm$100-200~m\,s$^{-1}$, reducing to $\sim$1-2~m\,s$^{-1}$ in integrated sunlight.  The difference between weak and strong lines is an order of magnitude smaller (Fig.\,\ref{fig:diff-diff}), perhaps $\sim$0.1-0.2~m\,s$^{-1}$ , requiring highly precise measurements for detection.  However, once such shifts between groups of different spectral lines can be reliably identified, that will constitute a specific signature (for the current discussion, at least of the non-magnetic) granulation, and suitably matched filtering should be able to identify and remove much of that particular atmospheric signature.

\begin{figure*}
\sidecaption
 \includegraphics[width=12.9cm]{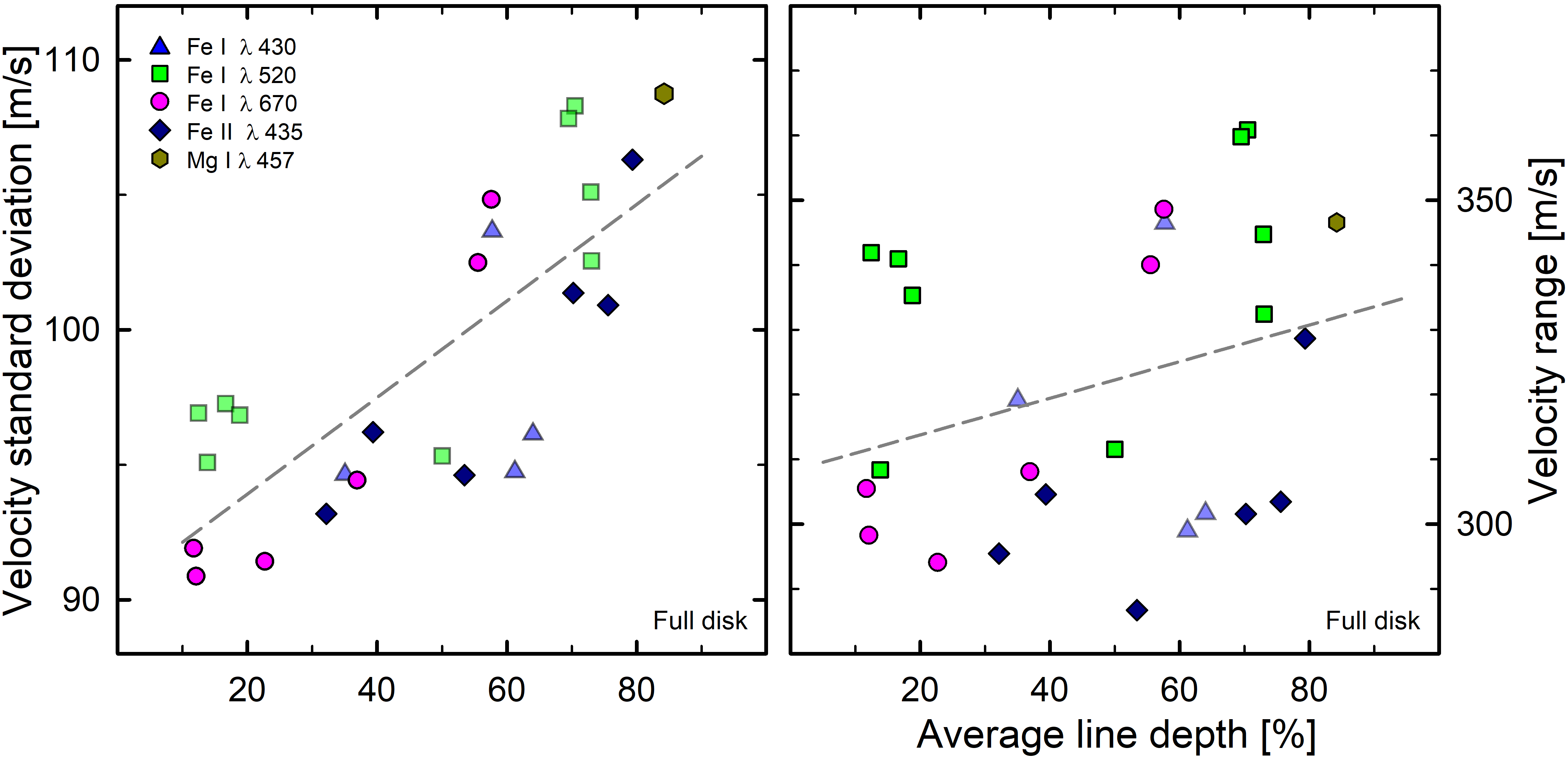}
     \caption{Radial-velocity fluctuations among different line-groups, as function of absorption-line depth.  Standard deviations and jittering ranges in the simulation of integrated sunlight are shown, with numbers from the small simulation area (for the actual Sun, divide by a factor of about 150-200).  Stronger and deeper lines display greater amplitudes.  The trends extend to the even stronger \ion{Na}{i} D$_{1}$ line which, at 136 and 404~m\,s$^{-1}$, is off scale and not included for the fitted lines. }  
\label{fig:jittering_amplitudes}
\end{figure*}

\subsection{Mechanism of radial-velocity differences}

For solar-type stars, the greatest convective blueshifts are found in weak lines formed in the deeper photosphere.  However, here we find that the variability of that quantity is smallest in those weak lines and instead increases for stronger lines predominantly formed in the upper photosphere.  The mechanism can be tentatively understood by examining a typical 3D atmospheric structure, e.g., Fig.\ 14  in \citet{nordlundetal09}.  That figure shows temperatures as a function of optical depth at several horizontal locations across a simulation area.  When plotted on an optical-depth scale (not versus geometric height), the temperature dependence with optical depth is rather similar in both hot granules, and in cool downflows. The reason is the steep temperature dependence of the opacity, and any given optical depth is reached at about the same temperature, irrespective of whether the temperature gradient is high (as in rising granules) or more modest (as in sinking downflows).  Local line strengths, however, depend also on the vertical gradients. 

As illustrated in that figure, the smallest temperature fluctuations occur at optical depths around $\tau$ = 0.3, reaching greater but comparable values both deeper down ($\tau$ = 1) and higher up, $\tau$\,=\,0.1.  Weaker spectral lines form at optical depths around about $\tau$ = 0.5, where the fluctuations between various horizontal locations are smaller than those on the more elevated formation heights of the stronger lines (perhaps $\tau$ = 0.1).  At such levels, the lines thus experience greater jittering amplitudes.  A velocity disturbance coming from below will propagate upward into layers of reduced optical depth and lower density, where the fluctuation amplitudes may roam more freely.  However, what matters is not only the excursion of the temperature as such, but rather how spectral-line parameters respond to temperature disturbances.  As seen in observed spectra of stars slightly hotter and slightly cooler than the Sun, temperature divergences from the solar value cause greater relative line-strength changes if the excursion is toward cooler temperatures (i.e., toward higher regions of stronger-line formation) than if it gets hotter; e.g., Fig.\ 4 in \citet{biazzoetal07}.  This holds for both \ion{Fe}{i} and \ion{Fe}{ii} lines, although their signs of change are opposite.  Such differential behavior is an established indicator of stellar temperature changes \citep{gray94}.

\subsection{Jittering amplitudes} 

Due to their smallness, some model predictions may not be easy to verify in current observations of the Sun seen-as-a-star.  However, the modeling could instead be tested in some detail with spatially resolved observations across the solar disk.  In Fig.\ \ref{fig:idealized_lines_rms}. the root-mean-square values for the velocity fluctuations are shown for idealized \ion{Fe}{i} lines at $\lambda$\,620 nm, for two line-strengths and two excitation potentials.     At the various center-to-limb positions ($\mu$ = cos\,$\theta$ = 1.00, 0.87, 0.59, 0.21), also the values for different azimuth angles are shown, together with their averages.  The multiple thin curves indicate the physical spread of 3D dynamics in the model and also serve to remind that averaging over rather extensive simulation sequences will be required to reliably identify also more subtle dependences or higher-order effects.

The jittering is clearly greatest at disk center, gradually decreasing toward the limb.  Over much of the disk (50\% of the disk area is at $\mu$\,$>$\,0.71), the stronger line displays the greatest jittering amplitude but closer to limb, the weak line starts to dominate.  Near disk center, the low-excitation lines jitter more but that reverses near the limb.  

Figure \ref{fig:jittering_amplitudes} instead shows jittering amplitudes for the full simulated solar disk, for different selected lines, now including the quite strong \ion{Mg}{i} $\lambda$\,457 nm.  Also, the even stronger \ion{Na}{i} D lines were measured.  Those will likely not be used for radial-velocity observations but their data serve to confirm trends seen already in weaker lines.  Figure \ref{fig:jittering_amplitudes} uses data from the full spectrum in the N59 model, and the velocity standard deviations refer to excursions around the mean velocity, as averaged over the simulation sequence. 

Figures \ref{fig:idealized_lines_rms} and \ref{fig:jittering_amplitudes} provide specific model predictions which should be possible to test observationally.  For measurements of integrated sunlight, the amplitudes from the simulation area have to be reduced by a large factor but only by a small amount (if any) in spatially resolved data.  Such observations should become practical with forthcoming solar telescopes, where light from sections of a spatially somewhat resolved solar image is fed into a high-precision instrument, e.g., the PoET project at the ESPRESSO spectrometer of ESO in Chile \citep{leiteetal22}.

 \begin{figure*}
 \sidecaption
 \includegraphics[width=12.9cm]{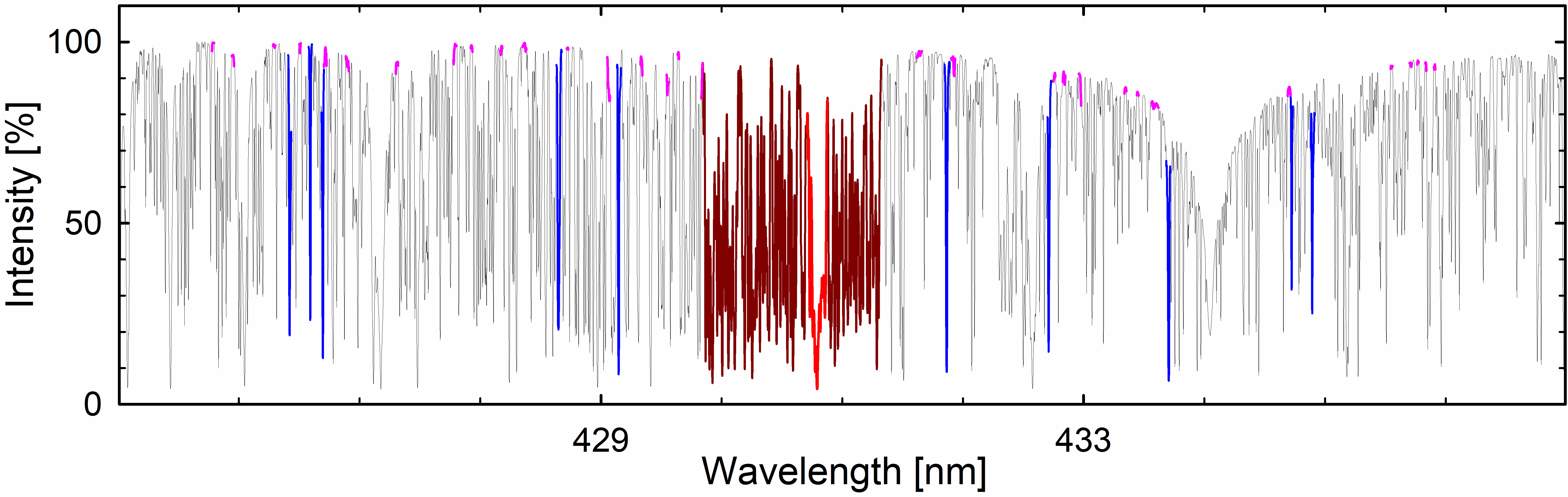}
     \caption{Spectral region surrounding the G-band, indicating measured sections: G-band core (red), full G-band (red + brown), nearby \ion{Fe}{i} lines (blue), and pseudocontinuum reference points (red). This plot from the CO\,$^5$BOLD N59 model is for disk center, $\mu$=1. }  
\label{fig:g-band-regions}  
\end{figure*}

\begin{figure*}
 \sidecaption
 \includegraphics[width=12.9cm]{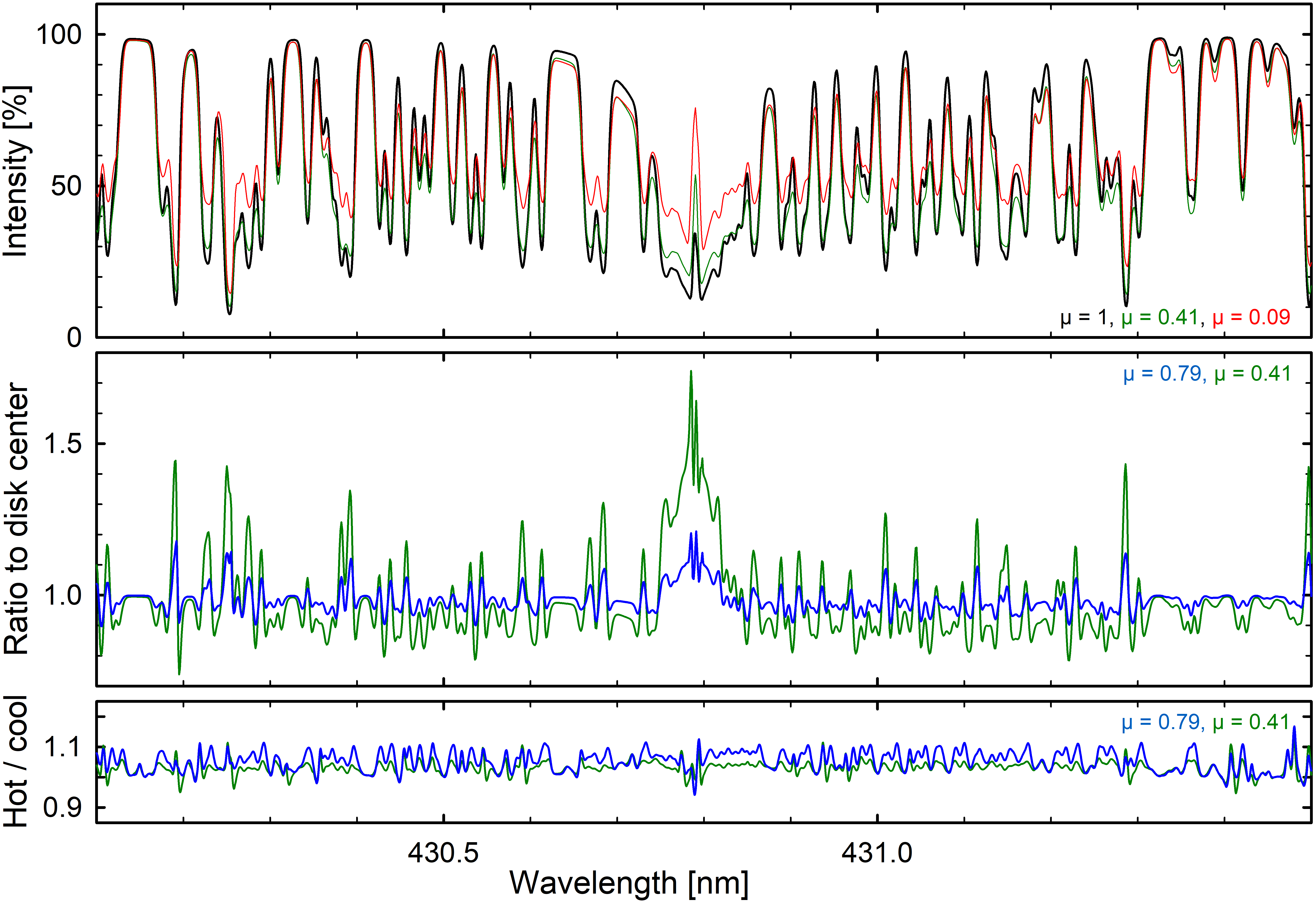}
     \caption{G-band region in synthetic non-magnetic CO\,$^5$BOLD N59 solar spectra.  Top frame: Intensity spectra at disk center $\mu$ = 1 (black), at $\mu$ = 0.41 (green), and close to limb, $\mu$ = 0.09 (red), separately normalized to a very local 'continuum`.  Middle frame: spectral ratios relative to disk center for $\mu$ = 0.79 (blue) and $\mu$ = 0.41 (green).  Bottom frame: Range of temporal variability is illustrated by spectral ratios between snapshot instances of highest (5806 K) and lowest (5735 K) model temperatures; same vertical scale as middle frame. }  
\label{fig:g-band_co5bold}
\end{figure*}

\section{Specific spectral regions}

There might exist spectral features that are both sensitive to fluctuations in the photospheric structure and also practical to observe as possible proxies for excursions in radial velocity.  Parameters of strong chromospheric lines have often been used to quantify solar activity, in particular the \ion{Ca}{ii}~H\,\&\,K emission, Balmer lines from hydrogen and high-excitation lines from helium. 

Nevertheless, all those have shortcomings in that they do not precisely reflect local conditions in the photosphere.  The strong chromospheric lines are formed mainly in high atmospheric layers with their greater amplitudes in gas motions, driven by magnetic structures, while the weaker He lines are partially excited by X-rays radiating downward from the transition region.  On the other hand, precision radial velocities are measured from photospheric spectra only, and one needs to find features that are particularly sensitive to photospheric changes and that also appear promising to examine in observed spectra.

\subsection{The G-band}
 
One interesting region is the G-band.  This molecular bandhead around 430 nm mainly consists of transitions between rotational and vibrational sublevels of the CH molecule, blended with numerous atomic lines.  The high line density causes an absence of any real continuum across the whole bandhead.  As discussed below, the detailed physics of its spectrum formation is not fully understood, although its formation must largely reflect photospheric structures.  Line identifications in that region are shown in the Sacramento Peak solar atlas \citep{beckersetal76}, in the IRSOL Second Solar Spectrum atlas \citep{gandorfer02}, and in the graphic version \citep{mrotzeketal16} of the Göttingen solar flux atlas \citep{reinersetal16}.  Here, we examine the theoretically expected types of variability in advance of actual measurements in Paper II.

\subsection{Non-magnetic G-band spectra}

Inside the G-band, we do not attempt to follow radial velocities since the massive blending makes it awkward to reliably segregate wavelength displacements from changes in line profiles induced by varying blends driven by, e.g., temperature fluctuations.  Instead, the total spectral flux within different regions is examined.  While the exact boundaries of the G-band are not uniquely defined, we adopt the full G-band as 429.84 -- 431.33 nm, with its central core 430.71 -- 430.88 nm (Fig.\ \ref{fig:g-band-regions}).  In its immediately surrounding region, ten reasonably clean \ion{Fe}{i} lines were selected to provide also radial-velocity measures: five lines shortward, and five longward of the G-band.  As a near-continuum reference, 32 segments were selected throughout the region (the same ones later used for observational studies in Paper II).  This enables to search for differing behavior among spectral features in this region. 

\begin{figure*}
 \sidecaption
 \includegraphics[width=12.9cm]{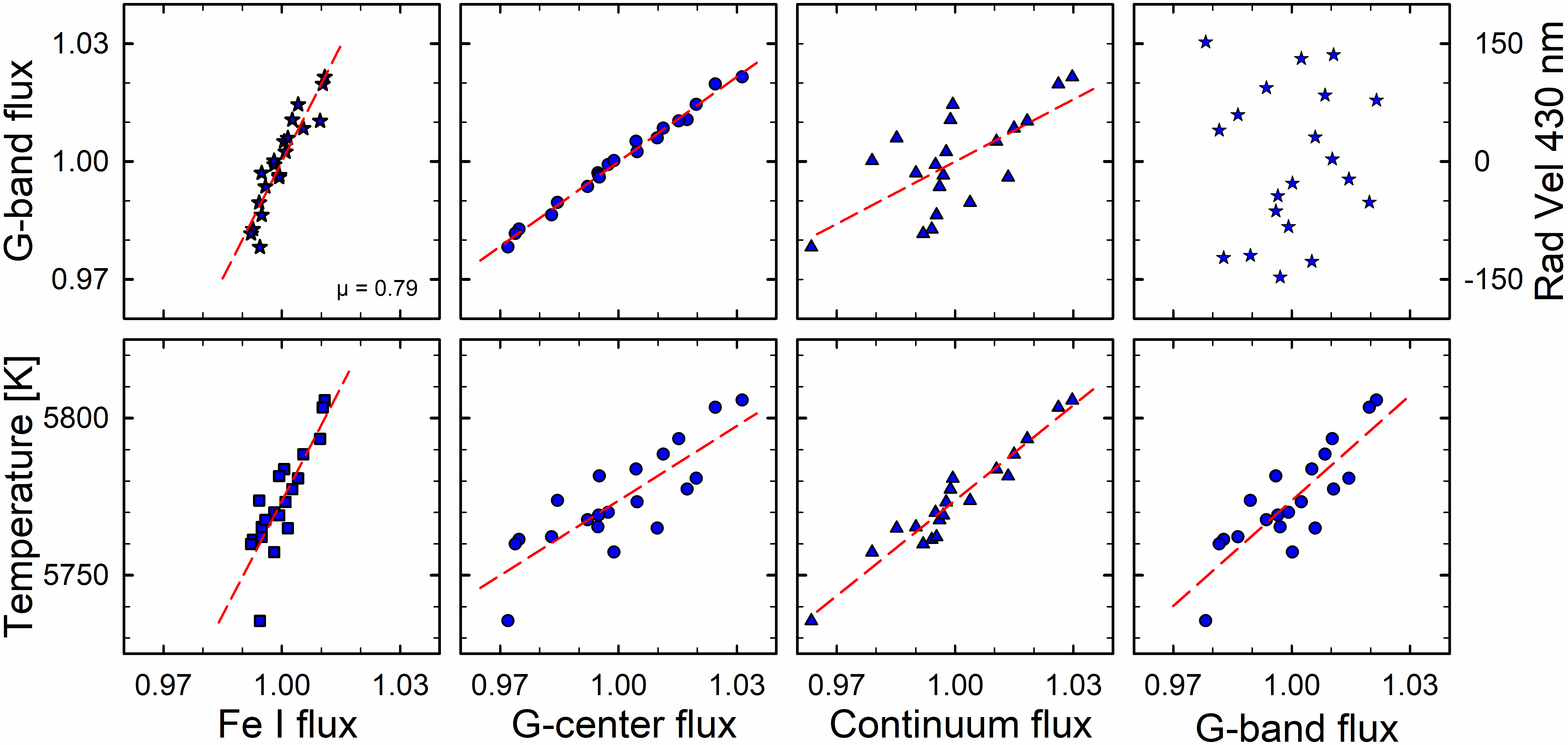}
     \caption{Temporal flickering of the flux in the G-band region of synthetic non-magnetic CO\,$^5$BOLD N59 solar spectra, measured at $\mu$ = 0.79.  The lower row shows how the effective temperature of the model area at each selected time of the simulation affects features in the region.  The G-band flux is plotted in the three leftmost upper frames; the rightmost one shows radial velocities for the full disk measured in the nearby \ion{Fe}{i} lines surrounding the G-band.  } 
\label{fig:g-band_correlations_b}
\end{figure*}

Synthetic G-band spectra at different center-to-limb locations are shown in Fig.\ \ref{fig:g-band_co5bold}, together with their ratios.  There is a striking dependence such that the relative brightness of the G-band core greatly increases when moving from disk center toward the limb.  By contrast, the changes induced by temperature fluctuations in the model are quite modest.  Spectral variability is thus more pronounced with respect to the spatial disk position, not with temporal changes.  Figure \ref{fig:g-band_correlations_b} shows the types of correlations present, here selected for $\mu$ = 0.79 as representative for much of the emerging flux from the full disk.  The data show integrated fluxes of each spectral feature inside its respective wavelength interval (the flux, not the absorption).  Naturally, fluxes increase during hotter instants of the simulation but the responses differ among different spectral features, with the G-band center showing greater spread than \ion{Fe}{i} lines.  However, the radial velocities (being the averages from the ten surrounding \ion{Fe}{i} lines) fail to show any obvious correlation with the G-band fluxes.

\subsection{Magnetic G-band spectra}

The complex molecular structure of the G-band implies that its formation may be influenced by numerous effects, as already suggested in the examples of Fig.\ \ref{fig:g-band_co5bold}.  A primary observation is that magnetic field concentrations in the quiet Sun outside sunspots are spatially largely coincident with areas that are bright in the G-band.  Such regions of magnetically affected granulation may cover a non-negligible fraction of the Sun and have a different structure and generally exhibit smaller convective wavelength shifts than non-magnetic areas.  Due to their area coverage, these facular and plage regions are the second most significant contributors to the emerging solar radiation (after the extensive regions of non-magnetic granulation).  Much of the solar-cycle variation in apparent solar radial velocity appears to be due to the varying area coverage of such magnetic plage regions \citep{meunieretal10a, meunieretal10b} and it would be desirable to identify some independent measure of their coverage.

\subsubsection{Solar filigree and their magnetic fields} 

A significant fraction of small-scale magnetic fields in the photosphere are traced by the elongated, sinuous and crinkly solar filigree, also referred to as bright points (although they seldom are point-like).  These small brightness enhancements on subarcsecond scales are now understood to form inside intergranular lanes, near the outer edges of granules, where a local minimum of gas pressure (contributing to the horizontal pressure gradient that drives horizontal gas motions inside and outside granules) is compensated by the internal magnetic pressure in the flux tubes to achieve a balance with their surroundings. 

Solar filigree were first observed (and named) by \citet{dunnzirker73}.  They could be resolved once subarcsecond angular resolution was achieved for imaging the photosphere in narrow-band monochromatic images, initially in the far wings of H$\alpha$ \citep{dravins74, dunnzirker73}, later at also other wavelengths and soon understood to be largely cospatial with magnetic field concentrations \citep{mehltretter74}.  The appearance of solar filigree in various spectral regions on the quiet Sun is described by \citet{bergertitle01, bergeretal95, bergeretal04, rouppevandervoortetal05} and \citet{sanchezalmeidaetal04}; a review is by \citet{bellotrubioorozvosuarez19}.  Among these, observations in the G-band are particularly extensive.

\begin{figure*}
\sidecaption
 \includegraphics[width=12.9cm]{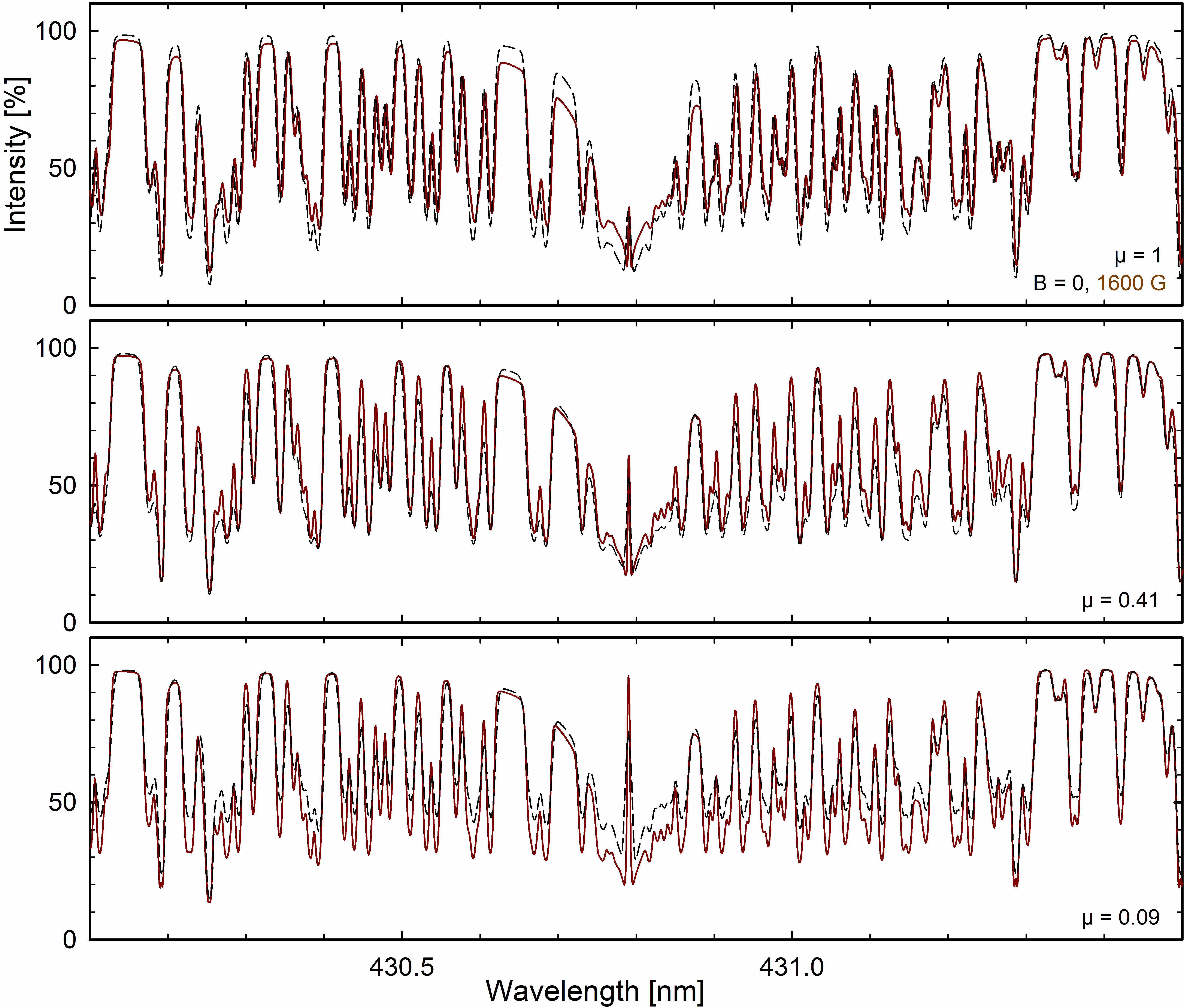}
     \caption{Region around the central G-band in  CO\,$^5$BOLD simulations for zero magnetic field (dashed black curves) and for B = 1600 G (160 mT); solid colored curves; for three different center-to-limb positions.  The spectra are individually normalized to their local pseudo-continua.}  
\label{fig:g-band_magnetic_a}
\end{figure*}

\subsubsection{Modeling G-band spectral brightness} 

Specifics about the G-band formation attracted attention already a long time ago when observations of its center-to-limb variation suggested it to be formed on a rough and corrugated solar surface \citep{pecker49}.  More recent work has mapped also the center-to-limb change in convective wavelength shift \citep{langhansschmidt02}.  Solar filigree are well visible in the G-band and numerous attempts have been made to model their brightness.  There is general agreement that the magnetic pressure inside flux tubes implies a reduced density relative to their surroundings.  At a given geometric height, a lower gas pressure causes a lower opacity than in adjacent non-magnetic regions, enabling to see deeper and hotter layers.  In these deep photospheric layers, the CH abundance is reduced by dissociation and its weaker absorption lines permit more continuum flux to shine through \citep{carlssonetal04, kiselmanetal01, shelyagetal04, schussleretal03, stein12, steineretal01, steineretal03}.

The detailed physics may be more complicated because of effects from non-LTE and three-dimensional radiative line transfer.  The hot walls of the flux tube perimeter emit enhanced ultraviolet radiation into the flux tube from its sides, driving photo-dissociation of molecules \citep{steineretal01}.  Since the G-band largely contains saturated absorption lines, the amount of its weakening is sensitive to specific line-broadening parameters \citep{sanchezalmeidaetal01}.  The appearance of the solar surface is further affected by its geometric corrugation: granules have their continuum optical depth unity level some 80 km above the photospheric average, with the magnetic flux tubes some 200–300 km below \citep{carlssonetal04}.  For those flux tubes that are inclined relative to the line of sight, the view toward deeper layers may be obstructed. 

It  should also be noted that not all magnetic elements necessarily are bright: In the quiet Sun, G-band brightenings almost always correspond to elements with strong magnetic field but in developed plages, the correlation becomes weaker \citep{riethmullersolanki17}.  Also, non-magnetic bright points may appear, at least in simulations, caused by vertically extending vortices (vortex tubes) in the photosphere \citep{calvoetal16}.

\subsubsection{Other molecular bands}

The low gas densities and high transparency of fluxtubes produce enhanced contrast in also other spectral features, in particular molecular lines. The violet CN band around 388 nm gives a contrast about 1.4 times greater than the G-band \citep{kiselmanetal01, ruttenetal01, zakharovetal05}, although observationally that short-wavelength region is more demanding than the G-band.  Responses for further molecular species are discussed by \citet{berdyuginaetal03, jietal16, steineretal01}, and \cite{tritschleruitenboek06}, that of extended spectra by \citet{norrisetal17, norrisetal23}.

\subsubsection{Magnetic  CO\,$^5$BOLD models} 

After injecting a certain magnetic flux in 3D atmospheric structures, its development can be followed into magnetic flux concentrations around intergranular lanes, the magnetic forces being balanced by gas pressure and dynamic effects.  The granular motions and their development is disturbed, resulting in a more disrupted granulation structure with a generally smaller convective blueshift.  Such surface magnetoconvection has been modeled, for example, by \citet{beecketal15a, beecketal15b}, \citet{khomenkoetal18}, \citet{norrisetal23}, and \citet{salhabetal18}.  Spectral synthesis illustrates how the magnetic properties are reflected in the shapes of spectral lines, depending on their magnetic sensitivity \citep{holzreutersolanki12, holzreutersolanki15}.   

Here we use a series of magnetic CO\,$^5$BOLD solar models \citep{ludwigsteffen22, ludwigetal23}.  The total net magnetic flux through each horizontal plane is kept constant with the field passing orthogonally through the lower and upper model boundaries, with flux densities of 800-1600-2400 G (80-160-240 mT).  This results in a somewhat deformed granulation structure, from which spectra of the G-band region were computed.  Values on the order of 160 mT seem characteristic for actual intergranular field strengths in solar filigree \citep{bellotrubioorozvosuarez19, dewijnetal09, salhabetal18, stein12}.  Figure \ref{fig:g-band_magnetic_a} compares the resulting spectra between the zero-field case and that of this representative intergranular flux density of 160 mT.  The overall spectral contrast varies with center-to-limb distance and changes sign when approaching the limb. 

\begin{figure}
 \centering
 \includegraphics[width=6cm]{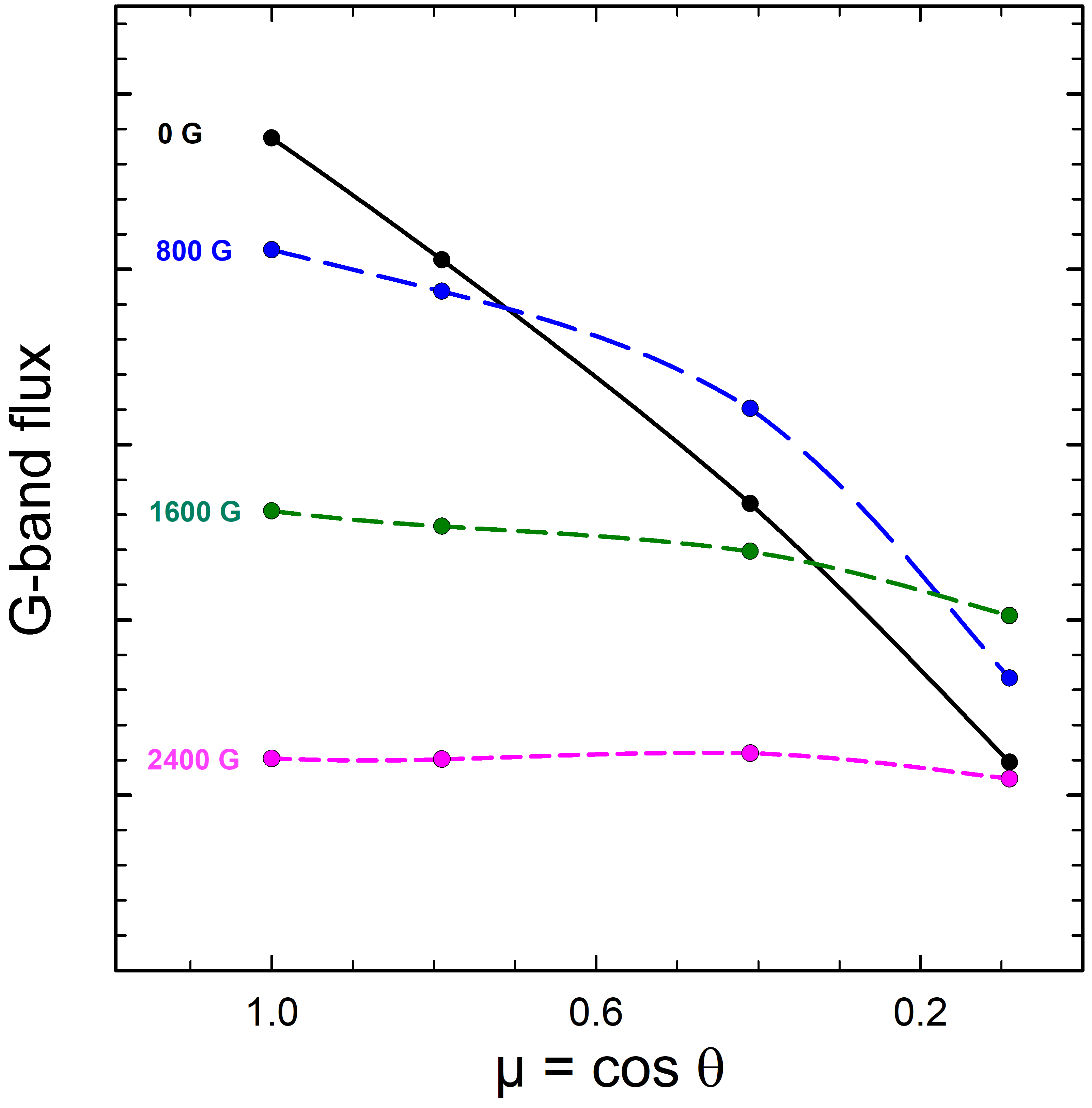}
     \caption{Emission levels in pseudo-continua around the G-band for different magnetic field strengths, relative to the zero-field case at solar disk center, and for different center-to-limb positions. For 800 and 1600\,G (80 and 160 mT), the crossover to bright regions (faculae) is seen when moving toward the limb.  For the strongest magnetic fields, the fluxes become faint; apparently dark pores begin to form. }  
\label{fig:g-band_magnetic_fluxes}
\end{figure}

This is further illustrated in Fig.\ \ref{fig:g-band_magnetic_fluxes}, which shows how the modeled brightness varies across the solar disk for different magnetic field strengths. The G-band flux is here given by the pseudo-continuum levels near the edges of the G-band region: 428.8 -- 428.9, and 431.9 -- 432.0 nm.  For moderate flux densities of 800 and 1600 G (80 and 160 mT), this continuum brightness overtakes the zero-field case in positions toward the limb, thus illustrating the crossover to bright facular regions, as distinctly observed in the G-band.  For the strongest fields treated here, the continuum intensity becomes faint throughout and apparently something akin to umbral dots begins to form.  With even stronger magnetic flux concentrations, surface convection is inhibited and starspots develop, forming umbrae and penumbrae structures, as modeled for different stellar types by \citet{panjaetal20}.  While outside the scope of the present paper, corresponding synthetic spectra from also such magnetic umbrae and penumbrae may be combined into a study of fluctuations of spatially resolved features.  Pairs of spectral lines with disparate Land{\'e} $\varg$$_{\textrm{eff}}$-factors and thus different magnetic sensitivities should be able to diagnose field properties \citep{smithasolanki17}, although it might be challenging to perform a detailed synthesis of a full spectrum that precisely incorporates all Zeeman components from all relevant lines.  Although the G-band is a very crowded region, there still exist instances of somewhat isolated groups of CH lines that are predicted to produce a measurable circular polarization signal through the Zeeman effect \citep{uitenbroeketal04}.

\section{Flickering in photometric colors}

We now examine what information might be extracted from fluctuations over even broader wavelength regions.  Precision measurements of stellar irradiance, made from space, reveal the brightness flickering due to time-variable granulation.  This has been theoretically modeled for different spectral types \citep{lundkvistetal21, rodriguezdiazetal22, samadietal13a,samadietal13b}, and its observation permits various stellar parameters to be inferred \citep{bastienetal16, nessetal18, vankootenetal21}.  Such photometric flicker is a concern for exoplanet transit photometry \citep{sulisetal20}.

The connection between simultaneous fluctuations in radial velocity and brightness has been studied.  Clear correlations are seen for hotter and more active stars, but are not identified for the smaller amplitudes in cooler stars of solar temperature \citep{giguereetal16, oshaghetal17, sulisetal23}.  In general, stars with greater photometric variability (probably caused by spots) also have increased levels of jitter in radial velocity.  Thus, photometric stability may be used as a criterion to select stars that remain particularly quiet in radial velocity \citep{hojjatpanahetal20, tayaretal19}.  

As discussed above, 3D simulations suggest relations where a 1\,m\,s$^{-1}$ fluctuation in velocity caused by granular evolution would correspond to a brightness fluctuation on the order of 100 ppm in the light from the full solar disk.  However, the longer-term photometric variability of the Sun is dominated by sunspots and active regions, even if searches for photometry as a proxy for short-term variations have been made \citep{kosiarekcrossfield20}.

The required micromagnitude accuracies in absolute stellar fluxes cannot realistically be achieved from the ground, but differential measures of photometric colors might perhaps be feasible from exceptional observing locations \citep[e.g.,][]{schmideretal22}, despite requiring a good understanding of atmospheric extinction, especially at the shorter wavelengths.  Color changes reflect stellar temperature variations and should carry information analogous to the flickering in brightness, possibly being a candidate for a proxy of excursions in radial velocity. Since our simulations provide absolute fluxes, synthetic broad-band photometric colors can be computed.   For CO\,$^5$BOLD models, certain synthetic colors have previously been computed by \citet{bonifacioetal18} and \citet{kucinskasetal18}, and for the STAGGER grid by \citet{chiavassaetal18}. 

A photometric system developed to give maximum sensitivity to physical parameters in specifically solar-type stars is the Str{\"o}mgren one with its four passbands of $u-v-b-y$, originally proposed by \citet{stromgren56, stromgren66}, with later calibrations by \citet{crawfordbarnes70}, \citet{onehagetal09}, and others.  In particular, it is being used to study properties of solar near-twins, e.g., \citet{melendezetal10} and \citet{dallemese20}.  

\begin{figure}
 \centering
 \includegraphics[width=\hsize]{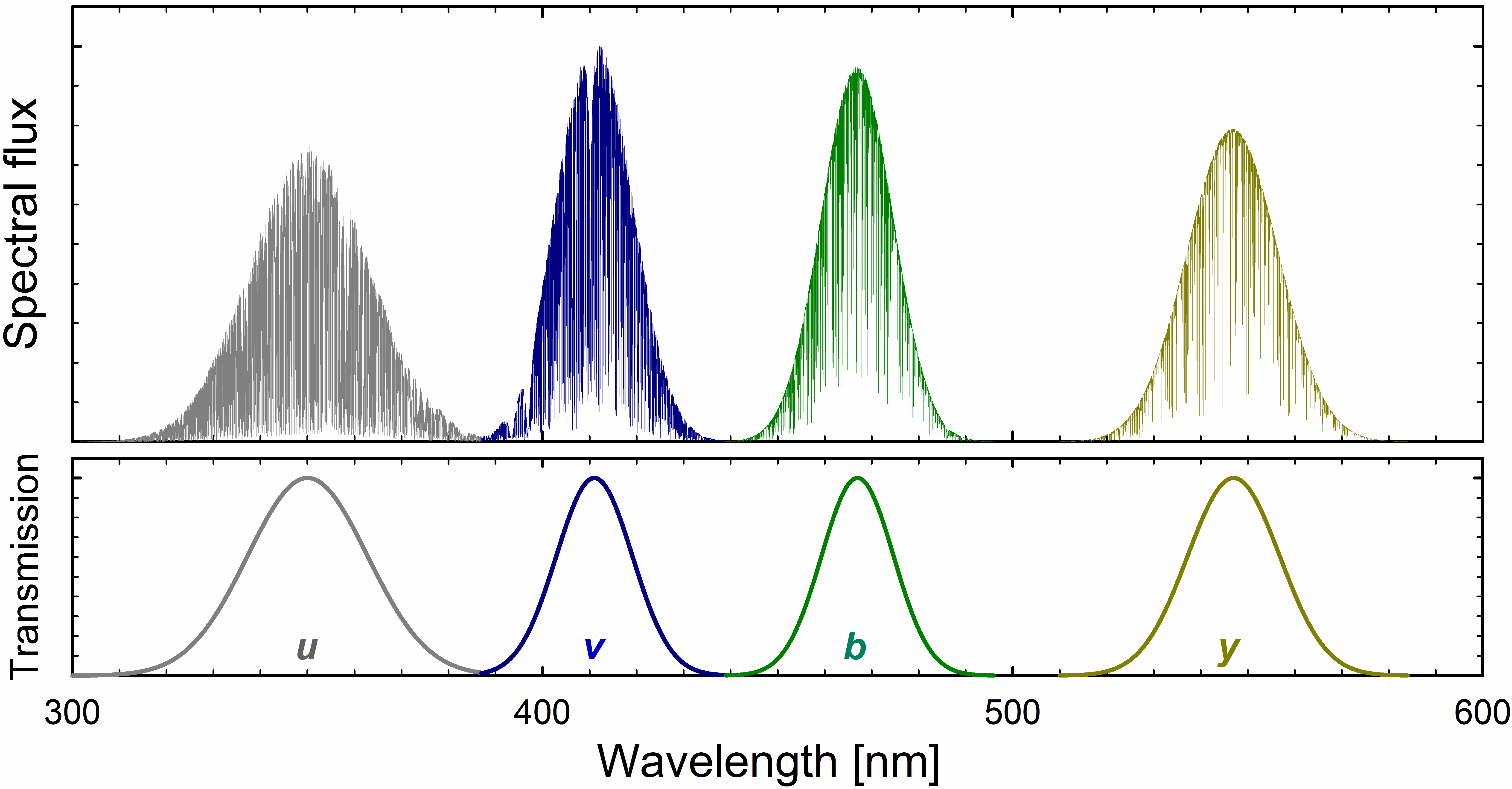}
     \caption{Top row: Spectral flux transmitted thorough the four Str{\"o}mgren $u-v-b-y$ filters, whose transmission functions are in the bottom row. }  
\label{fig:uvby-filters}
\end{figure}

These filters are defined such that the $y$ and $b$ passbands (centered around 547 and 467 nm) contain no strong spectral features for solar-type type stars, and thus the slope of the spectrum, measured as the color index $(b-y)$, is  free of significant effects from line blanketing, becoming an indicator of stellar temperature.  The $v$ filter, centered at 411 nm, is placed longward from the region where the Balmer lines from hydrogen begin to crowd, as those start to approach the Balmer limit (although the filter transmits H$\delta$).  The near-ultraviolet $u$ filter is centered at 350 nm with its longward part remaining below the Balmer jump while its shortward part extends to the telluric atmospheric absorption cutoff (Fig.\ref{fig:uvby-filters}).   Following standard usage, the fluxes $F$ inside each photometric $u-v-b-y$ band are converted to magnitudes according to $m= -2.5\,{\log}{_{10}}(F/F{_0})$, where $F{_0}$ is some reference flux.  However, here we are concerned only with the temporal fluctuations in color indices, not their absolute calibration.  For an overview of photometric systems, see \citet{bessell05}.  

The exact realization of observational $uvby$ photometry involves nuances such as the construction of the transmission filter optics, properties of the detector (e.g., whether integrating the energy flux inside the filter passband or the number of photons; here the former is applied).  We approximate the four filter passbands with Gaussians centered on 350, 411, 467, and 547 nm, with respective standard deviations 12.74, 8.07, 7.64, and 9.77 nm.  At the lowest transmission, they are truncated at 0.001 where adjacent filters would start to overlap at filter edges.  These profiles closely approximate those used in other work, and are shown in Figure \ref{fig:uvby-filters} together with the transmitted flux from the N59 solar model.

From measurements in these filter passbands, three indices are widely used for stellar classification: the color index $(b-y)$, and the two color-index differences $m{_1} \equiv (v-b)-(b-y)$, measuring the amount of absorption-line blanketing in the 410 nm region, and $c{_1} \equiv (u-v)-(v-b)$, quantifying the pressure-dependent Balmer discontinuity.  The color index $(b-y)$ reflects stellar temperature and the metal-line index $m{_1}$ indicates the amount of metallic-line absorption.  The Balmer discontinuity index $c{_1}$ provides a measure of the atmospheric pressure, which, being a function of the surface gravity, gives the stellar luminosity.  Greater values of $(b-y)$ imply cooler temperatures; increased metallicity is reflected in a larger $m_1$, and greater luminosity (lower pressure) as a greater value of $c_1$.  During the solar activity cycle, spectral brightness changes are driven by magnetic activity, influencing the signals in especially the Str{\"o}mgren $v$ and $b$ filters \citep{witzkeetal18}.    

\begin{figure}
 \centering
 \includegraphics[width=\hsize]{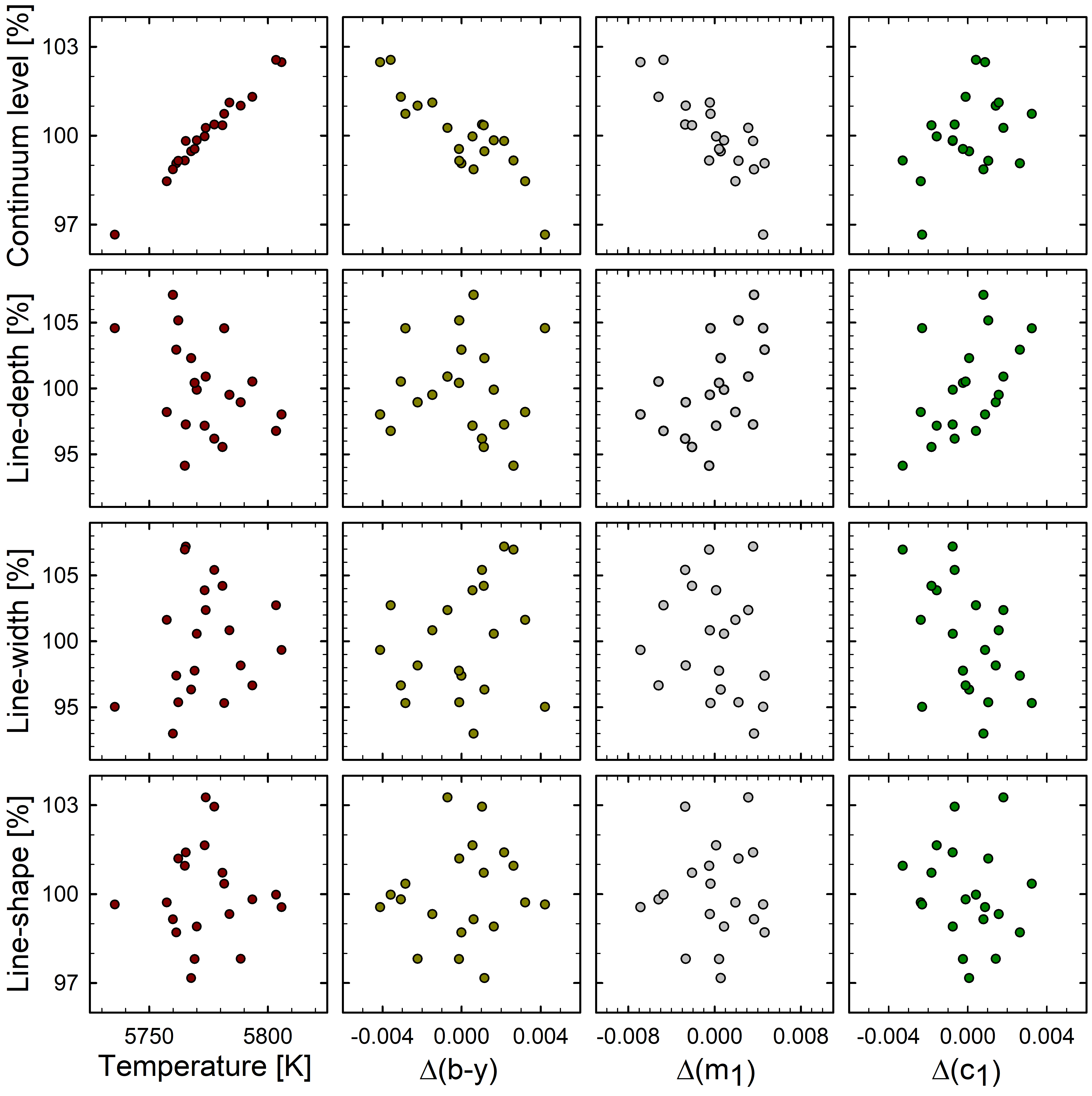}
     \caption{Relations between parameters of a representative \ion{Fe}{i} line of medium strength at 520 nm (Fig.\ref{fig:co5bold_525nm}) in integrated sunlight from the N59 model, and flickering in photometric Str{\"o}mgren indices.  Quantitites refer to relative variations over the small simulation area. }  
\label{fig:uvbycolors_a}
\end{figure}

\begin{figure}
 \centering
 \includegraphics[width=\hsize]{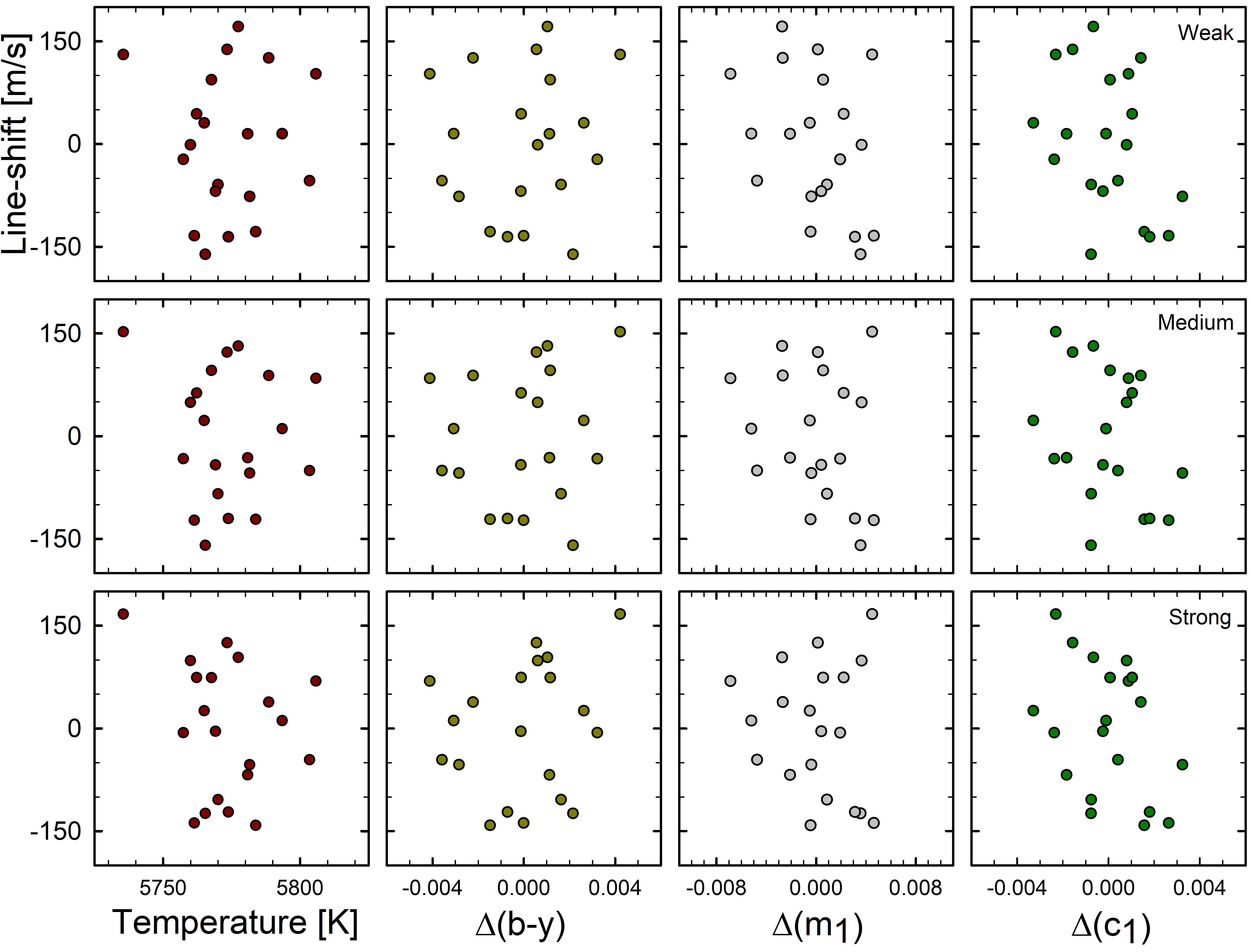}
     \caption{Search for relations between radial velocities of \ion{Fe}{i} lines of different strengths at 520 nm (Fig.\ref{fig:co5bold_525nm}), and photometric Str{\"o}mgren indices in integrated sunlight, as computed over the small simulation area of the N59 model (for the actual Sun, divide the numerical values by a factor $\sim$150-200).  Rows from top down: Weak, medium, and strong lines.}  
\label{fig:uvbycolors_b}
\end{figure}

Figures \ref{fig:uvbycolors_a} and \ref{fig:uvbycolors_b} show relations between model fluctuations in the various $uvby$ indices, also relating to the simultaneous jittering in apparent radial velocities for representative \ion{Fe}{i} lines of different strength.  Their parameters of continuum level, depth, width, shape and wavelength position are those obtained from the five-parameter Gaussian-type fit described above.  Clear trends are that instances of enhanced continuum brightening correlate with a bluer $(b-y)$ color, weaker line absorption, and an apparently lower metallicity.  However, relations among other measures of line-depth, line-width and line-shape are less obvious or absent, consistent with observational findings by \citet{oshaghetal17} and \citet{sulisetal23}.  

Irrespective of whether correlations with specific spectral signatures can be identified, the data illustrate what amplitudes might be expected in the flickering of full-disk sunlight in different color indices.  For the data shown in Fig.\ \ref{fig:uvbycolors_a}, the RMS fluctuations in $(b-y)$, $m{_1}$, and $c{_1}$ are, respectively, 0.0023, 0.0033, and 0.0017.  The greatest changes occur in $m{_1}$, due to changing absorption-line strengths in the blue while the modulation of the luminosity index is only half of that.  In actual full-disk sunlight, these amplitudes should be divided by a factor $\sim$150 or 200, then shrinking from a few millimagnitudes to only some 10 or 20 $\mu$mag.  The corresponding RMS fluctuations in the continuum level of 1.4\% at 525 nm over the model area correspond to some 100 ppm in full integrated sunlight, in line with what was deduced by \citet{ceglaetal19}.  Plots for separate center-to-limb positions show the relative fluctuation amplitudes increasing from disk center toward the limb but are not qualitatively different. The expected performance of space missions such as PLATO readily permits to detect such levels of photometric variability \citep{helleretal22,morrisetal20}, although their search for exoplanets will depend also on how to segregate other types of stellar variability.  Observed day-to-day solar irradiance fluctuations, driven by the presence of faculae and appearance of spots, are typically greater by an order of magnitude, and those during a solar activity cycle by another order of magnitude \citep{leanetal20, marchenkoetal19}.

\section{Lack of correlation in other line parameters}

\begin{figure}
\centering
 \includegraphics[width=\hsize]{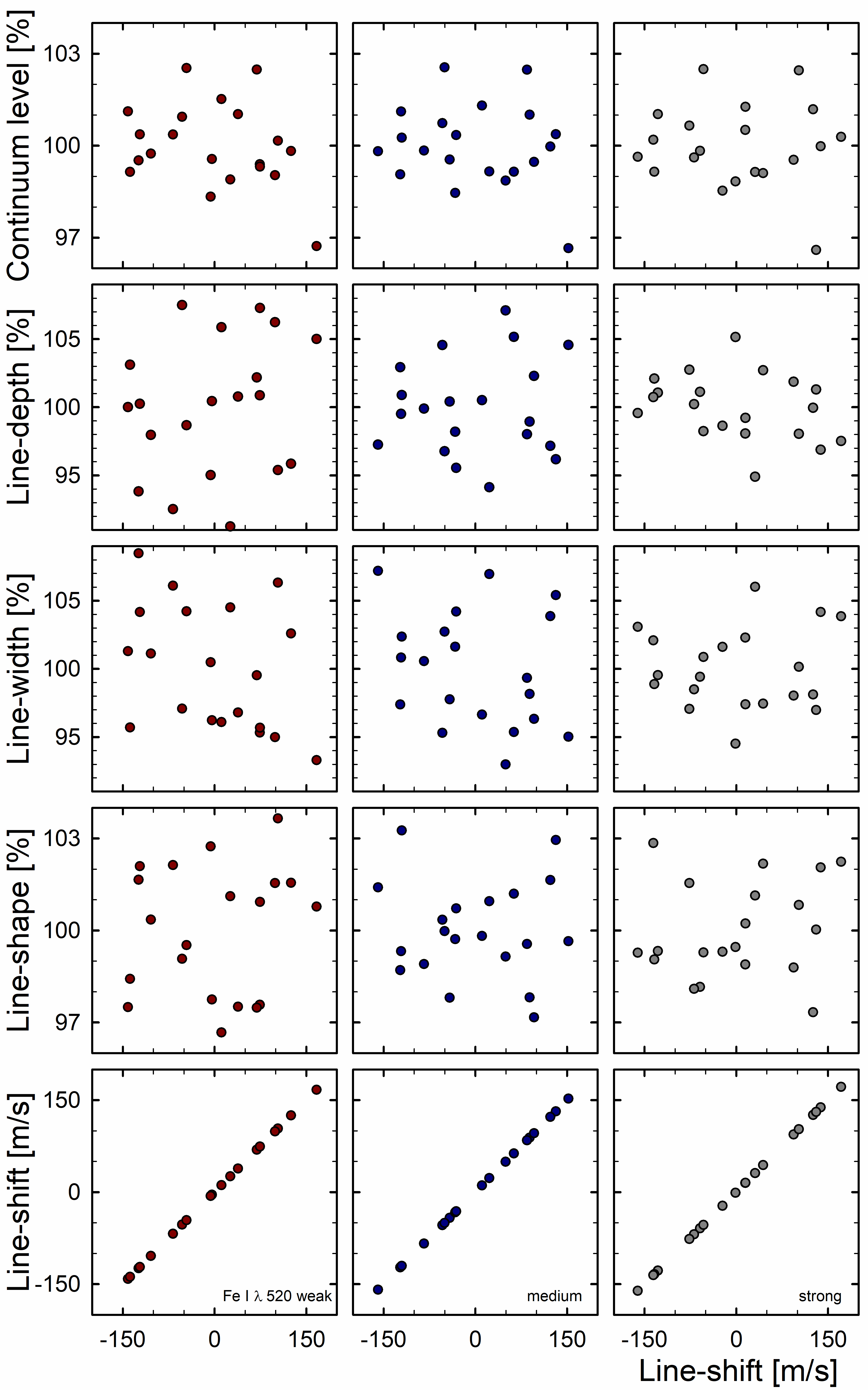}
     \caption{Radial velocities show lack of correlation with several plausibly anticipated fluctuations in various line parameters.  No sensible correlations were found here, except a hint for a line-depth dependence in the strongest lines.  The measures for line depth, width, shape are normalized to 100\% for the averages over the simulation sequence.  The calculations are for integrated sunlight from the small simulation area of the N59 model (for the actual Sun, divide the line-shift values by a factor $\sim$150-200). }  
\label{fig:no_correlations}
\end{figure}

Relations were sought between the velocity jittering and other spectral-line parameters which could have been suspected to vary in phase during the simulation sequences.  However, as seen in Fig.\ \ref{fig:no_correlations}, there appears to be no sensible correlation with either continuum brightness, line-width or line shape (the quantities in the five-parameter Gaussian-type fit described in Sect.\ 4.1).  An exception is that a weak relation appears to exist with line-depth, at least for the strongest lines.  Such a correlation actually is expected and was seen in clean disk-center data by \citet{dravinsetal21b}.  Locations of increased irradiance display enhanced convective blueshift combined with a stronger line absorption, caused by enhanced vertical gradients in regions of line formation.  However, while this is seen when granules are viewed directly from above, the effect rapidly diminishes away from disk center and largely gets washed out in the integrated sunlight of Fig.\ \ref{fig:no_correlations}.

\section{Search for Earth-like exoplanets}

Under realistic conditions, the spectral-line parameter that can be most precisely measured is its wavelength (rather than the depth, width or shape, which are more sensitive to the exact adjustments in the spectroscopic instrumentation).  Given the results in this paper, that is satisfactory since significant mutual correlations between line-groups could be identified for precisely such wavelength jittering.  Various other parameters, such as the fluctuating widths or depths of absorption lines, or flickering in broad-band photometric indices fail to show convincing correlations with excursions in radial velocity. 

Using the results from hydrodynamic simulations, it is thus possible to identify patterns of velocity jittering.  Characteristics of simultaneous jittering in radial velocity between different groups of lines, as shown in Figs.\ \ref{fig:diff_shifts_520nm} and \ref{fig:diff-diff}, discern the radial-velocity contributions originating from fluctuations in photospheric granulation, the signature of which should be possible to deduce and deduct. 

In practice, this will require to obtain radial-velocity measures for groups of individually selected spectral lines within distinct batches of stronger and weaker lines, and in different wavelength regions.  Jittering amplitudes from each such homogeneous line-group are expected to show systematic differences, e.g., such that the excursions of strong short-wavelength \ion{Fe}{ii} lines would be greater by a certain factor than, e.g., those of weak \ion{Fe}{i} lines at long-wavelengths.  A `matched filter' fitted to such a variability pattern should identify that component of velocity jittering which is specific to granular convection.  While the present work permits to estimate the rough parameters for such a filter, its more precise numbers will require more extensive simulations, the validity of which could be confirmed in such spatially resolved observations that are feasible already with current instrumentation, as suggested in Figs.\ \ref{fig:idealized_lines_rms} and \ref{fig:jittering_amplitudes}.  Of course, other sources of jittering originate from oscillations and various magnetic phenomena over other timescales, requiring separate attention.

The observational requirements are stringent but probably not impossible.  The total jittering amplitude in full-disk sunlight amounts to $\sim$1-2~m\,s$^{-1}$, with differences between various line-groups perhaps one tenth of that.  To mitigate these fluctuations thus requires to recognize the characteristic patterns of dissimilar lineshifts among classes of various spectral lines with precisions $\sim$10~cm\,s$^{-1}$ for each absorption-line group.  Although certainly challenging, such requirements correspond to what in any case is required in searches for truly Earth-like exoplanets; e.g., HARPS-3 \citep{farretjentinketal22, thompsonetal16} or ANDES \citep{marconietal22}. 

From other considerations, it has also been recognized that there is additional information content in individual spectral lines, as compared to statistical correlation functions \citep{almoullaetal22, cretignieretal20, dumusque18, meunieretal17b, wiseetal22}.  Recent efforts for the Sun seen as a star include the extraction of radial-velocity data separately from lines with different depths \citep{miklosetal20}, and the study of how spectral portions differ during epochs of magnetically more or less active periods \citep{thompsonetal20}. 

Most of our present modeling is limited to the non-magnetic photosphere and to such short-term fluctuations that are comparable to granular evolution timescales.  This quiet photosphere covers the largest fraction of the solar surface and contributes the greatest part of the total flux (and presumably that of other solar-type stars) but magnetic features seem responsible for much of the variability over the longer timescales of stellar rotation and activity cycles. 

The following Paper II \citep{dravinsludwig24} will explore observed solar microvariability on also such somewhat longer timescales, aiming to identify further measurable photospheric parameters that might be used as proxies for radial-velocity fluctuations.

\begin{acknowledgements}
{The work by DD is supported by grants from The Royal Physiographic Society of Lund.  HGL gratefully acknowledges financial support by the Deutsche Forschungsgemeinschaft (DFG, German Research Foundation) -- Project ID~138713538--SFB~881 (`The Milky Way System', subproject A04).  Extensive use was made of NASA’s ADS Bibliographic Services and the arXiv$^{\circledR}$ distribution service. We thank the referee for very detailed and insightful comments.}

\end{acknowledgements}


\end{document}